\documentclass[%
aip,jap,
 amsmath,amssymb,floatfix,
 preprint,%
]{revtex4-1}

\usepackage[utf8]{inputenc}
\usepackage[T1]{fontenc}
\usepackage{array}
\usepackage{amsmath,amssymb,amsfonts}
\usepackage{algorithmicx}
\usepackage{mathtools}
\usepackage{algorithm}
\usepackage{algpseudocode}
\usepackage{graphicx}
\usepackage{textcomp}
\usepackage{xcolor}
\usepackage{siunitx}
\usepackage{mathtools}
\DeclarePairedDelimiter\abs{\lvert}{\rvert}%

\def\BibTeX{{\rm B\kern-.05em{\sc i\kern-.025em b}\kern-.08em
    T\kern-.1667em\lower.7ex\hbox{E}\kern-.125emX}}
\begin{document}

\title{Multi-frequency Data Parallel Spin Wave Logic Gates
}

\author{Abdulqader Mahmoud}
\email{a.n.n.mahmoud@tudelft.nl}
\affiliation{Delft University of Technology, Department of Quantum and Computer Engineering, 2628 CD Delft, The Netherlands}

\author{Frederic Vanderveken}
\affiliation{KU Leuven, Department of Materials, SIEM, 3001 Leuven, Belgium}
\affiliation{Imec, 3001 Leuven, Belgium}

\author{Christoph Adelmann}
\affiliation{Imec, 3001 Leuven, Belgium}

\author{Florin Ciubotaru}
\affiliation{Imec, 3001 Leuven, Belgium}

\author{Said Hamdioui}
\affiliation{Delft University of Technology, Department of Quantum and Computer Engineering, 2628 CD Delft, The Netherlands}

\author{Sorin Cotofana}
\email{S.D.Cotofana@tudelft.nl}
\affiliation{Delft University of Technology, Department of Quantum and Computer Engineering, 2628 CD Delft, The Netherlands}

\begin{abstract}
By their very nature, Spin Waves (SWs) with different frequencies can propagate through the same waveguide without affecting each other, while only interfering with their own species. Therefore, more SW encoded data sets can coexist, propagate, and interact in parallel, which opens the road towards hardware replication free parallel data processing. In this paper, we take advantage of these features and propose a novel data parallel spin wave based computing approach. To explain and validate the proposed concept, byte-wide $2$-input XOR and $3$-input Majority gates are implemented and validated by means of Object Oriented MicroMagnetic Framework (OOMMF) simulations. Furthermore, we introduce an optimization algorithm meant to minimize the area overhead associated with multifrequency operation and demonstrate that it diminishes the byte-wide gate area by $30$\% and $41$\% for XOR and Majority implementations, respectively. To get inside on the practical implications of our proposal we compare the byte-wide gates with conventional functionally equivalent scalar SW gate based implementations in terms of area, delay, and power consumption.  Our results indicate that the area optimized $8$-bit $2$-input XOR and $3$-input Majority gates require $4.47$x and $4.16$x less area, respectively, at the expense of $5$\% and $7$\% delay increase, respectively, without inducing any power consumption overhead. Finally, we discuss factors that are limiting the currently achievable parallelism to $8$ for phase based gate output detection and demonstrate by means of OOMMF simulations that this can be increased $16$ for threshold based detection based gates.
\end{abstract}

\maketitle

\section{Introduction}
The amount of row data has rapidly increased in the last few decades due to the information technology unprecedented growth. These data are usually processed on high efficiency CMOS technology based computing platforms \cite{data1,data2,ITRS} and as the amount of row data increased, technology feature size has been shrunken to keep up with the computation power demands. However, when entering into the deca-nanometer regime CMOS downscaling becomes more difficult due to: (i) leakage wall \cite{cmosscaling2,cmosscaling3}, (ii) reliability wall \cite{cmosscaling1}, and (iii) cost wall  \cite{cmosscaling1,cmosscaling2}, which suggests the near end of Moore's law. As a result, different technologies, e.g., graphene  \cite{Yande1,Yande2,graphine1,graphine2,graphine3}, memristor \cite{memristor7,memristor8,memristor9,memristor10,memristor11}, spintronics \cite{spintronics6,spintronics7,spintronics8,spintronics9,spintronics10} have been explored in an attempt to meet the exponentially increasing computing market  demands \cite{survey2}. 

While each of these alternative technologies exhibits both strong and weak points, spintronics on its Spin Wave (SW) flavour seems to have a great potential to meet market needs \cite{survey2} due to its: (i) Ultra-low power consumption as no charge movements are required in order to perform calculations, (ii) acceptable delay, (iii) down to \SI{}{nm} range scalability, and (iv) natural support for data  parallelism enabled by the fact that SWs of different frequency can coexist and selectively interact within the same waveguide. 

In view of this, different logic gates built on spin wave technology were presented, e.g., \cite{logic21,logic12,logic11,logic17,logic25,logic4,logic16,logic18,logic24,Magnon_transistor, logic1, logic13,logic14,logic20,logic19,logic2,logic3,logic100,logic101,parallel_data_processing1}, and in the sequel we briefly present some of them. A current controlled Macha-Zender interferometer based NOT gate has been the first experimentally demonstrated SW logic gate \cite{logic21} and by making use of a similar method, other logic gates including XNOR, NAND, and NOR were realized \cite{logic12,logic11,logic17}. NOT, OR, and AND gates were designed using three terminal devices with transmission lines \cite{logic25}\cite{logic4}\cite{logic16}\cite{logic18} and voltage-controlled XNOR and NAND gates utilizing re-configurable nano-channel magnonic devices were suggested \cite{logic24}. In addition, an XOR gate was proposed by embedding magnon transistors between the Mach-Zehnder interferometer arms \cite{Magnon_transistor}. 
By relying on another information encoding method, i.e., on  SW phase rather than on SW amplitude as it is the case for the previously mentioned schemes, buffer, NOT, (N)AND, (N)OR, XOR, and Majority gates were introduced in \cite{logic1}. Moreover, alternative Majority gate designs were suggested to decrease the SW back propagation and increase the SW transmission efficiency \cite{logic13,logic14,logic20}. OR and NOR gates were designed using cross structures \cite{logic19} and physically implemented Majority gates were reported in \cite{logic2,logic3,logic100,logic101}. 

All the previously mentioned designs operate on same frequency SWs, i.e., on $1$-bit inputs, therefore, if multiple-bit input  functions are to be evaluated, e.g., bitwise XOR over two $n$-bit inputs $A=(a_1, a_2, \ldots, a_n)$ and $B=(b_1, b_2, \ldots, b_n)$, an XOR gate structure must be replicated $n$  times in order to process the $n$ input bit-pairs (sets) in parallel at the expense of area overhead.  However, different frequency SWs can simultaneously propagate through the same waveguide without affecting each other, while only interfering with their own species. This suggests that if each input pair $(a_i,b_i)$ is encoded with $f_i$ frequency SWs, XOR($A,B$) can be potentially evaluated with one instead of $n$ XOR gates.  This approach has been pursued in \cite{parallel_data_processing1}, which introduces a Majority gate structure able to simultaneously  process $3$ data set encoded at $3$ different frequencies. However, the suggested structure make use of bent regions, which have detrimental effects on SW propagation, and contains a magnonic crystal that induces a large delay overhead. 

In this paper we revisit the SW parallelism concept and propose a novel multi-frequency data parallel in-line generic SW gate structure. Our contributions can be summarized as follows:\\
\begin{itemize}
\item Generic multi-frequency data parallel in-line SW gate structure and an associated area optimization algorithm.
\item Design and validation of $8$-bit data parallel in-line Spin Wave logic gates: $8$-bit $3$-input Majority and $2$-input XOR gates are instantiated and validated by means of Object Oriented MicroMagnetic Framework (OOMMF) simulations.
\item Performance assessment and comparison with SW state-of-the-art: The proposed $8$-bit $3$-input Majority and $2$-input XOR gates require $4.47$x and $4.16$x less area, respectively, when compared with functionally equivalent scalar SW gate based implementations, at the expense of $5$\% and $7$\% delay penalty, respectively, and no power consumption overhead.
\item Parallelism limit study:  Demonstrate by means of OOMMF simulation that the maximum currently achievable parallelism, i.e., the number of different SW frequencies, is $8$ for phase based output detection and $16$ when spin wave magnetization is utilized to detect the gate output.
\item Design and OOMMF validation of a $16$-bit data parallel in-line Spin Wave $2$-input XOR gate.
\end{itemize}

The reminder of the paper is organized as follows. Section \ref{sec:Basics and background of SW technology} briefly explains the SW physics fundamentals and the associated computing paradigm. Section \ref{sec:Proposed Parallelism Structure} describes  the proposed $n$-bit data parallel SW logic gate and introduces the associated area optimization algorithm. Section \ref{sec:Simulation setup and experiments} provides inside on the utilized simulation platform and parameters, and presents simulation expperiments related to the validation of the $8$-bit $3$-input Majority and $2$-input XOR gates. Section \ref{sec:Simulation results and discussion} presents evaluation results for the two byte wide parallel gates and a comparison with functional equivalent scalar implementations. In addition, it discusses fan-in and geometric scalability, and maximum achievable parallelism issues, and variability and thermal noise effects. Section \ref{sec:Conclusion} concludes the paper.

\section{SW Based Computing Background}
\label{sec:Basics and background of SW technology}
When a ferromagnetic material is exposed to an external magnetic field  electron spins arrange themselves in the applied magnetic field direction,  in order to bring the total system energy to the lowest possible level \cite{Magnonic_crystals_for_data_processing}. Further, if the electron spins are deflected by an excitation method, e.g., by means of Magnetoelectric (ME) cell, antenna, a Spin Wave (SW) is created mainly due to exchange and dipole spin interactions. The precessional electron spin movement \cite{Magnonic_crystals_for_data_processing}, can be described by the Landau-Lifshitz-Gilbert (LLG) relation \cite{LL_eq,G_eq} as follows:
\begin{equation} \label{eq:1}
\frac{d\vec{m}}{dt} =-\abs{\gamma} \mu_0 \left (\vec{m} \times \vec{H}_{eff} \right ) + \alpha \left (\vec{m} \times \frac{d\vec{m}}{dt}\right ),
\end{equation}
where $\gamma$ is the gyromagnetic ratio, $\mu_0$ the vacuum permeability, $\alpha$ the damping factor, $m$ the magnetization, and $H_{eff}$ the effective field and it is expressed as:
\begin{equation} \label{eq:2}
H_{eff}=H_{ext}+H_{ex}+H_{demag}+H_{ani},
\end{equation}
where $H_{ext}$ is the external field, $H_{ex}$ the exchange field, $H_{demag}$ the demagnetizing field, and $H_{ani}$ the magneto-crystalline anisotropy.

\begin{figure}[t]
\centering
  \includegraphics[width=\linewidth]{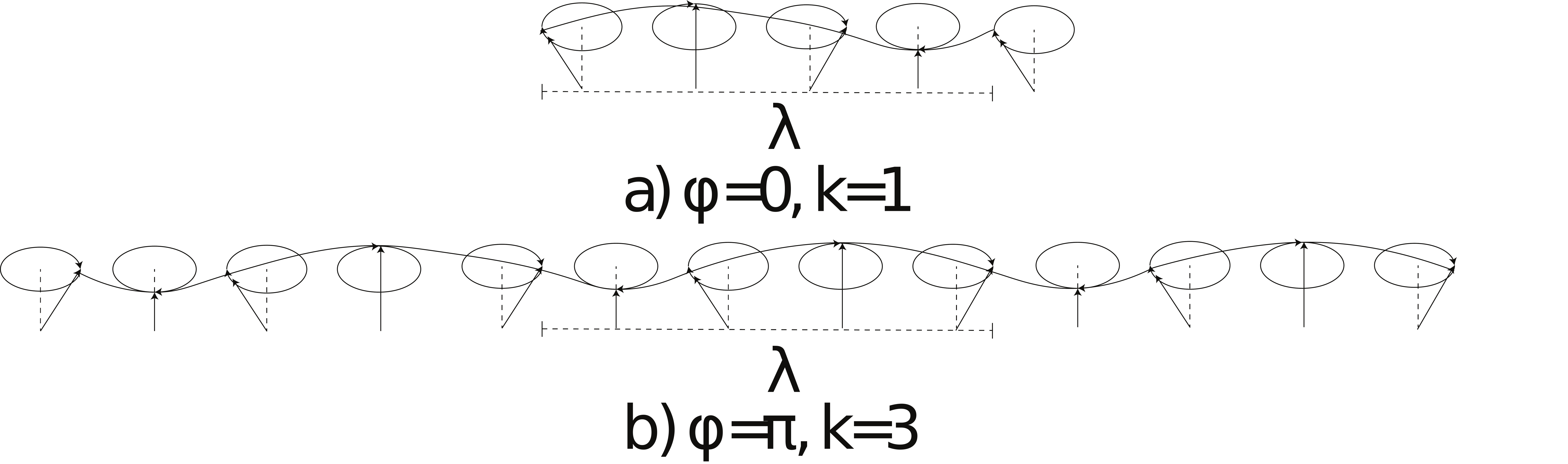}
  \caption{SW Parameters}
  \label{fig:SW_parameters}
\end{figure} 

\begin{figure}
\centering
  \includegraphics[width=0.6\linewidth]{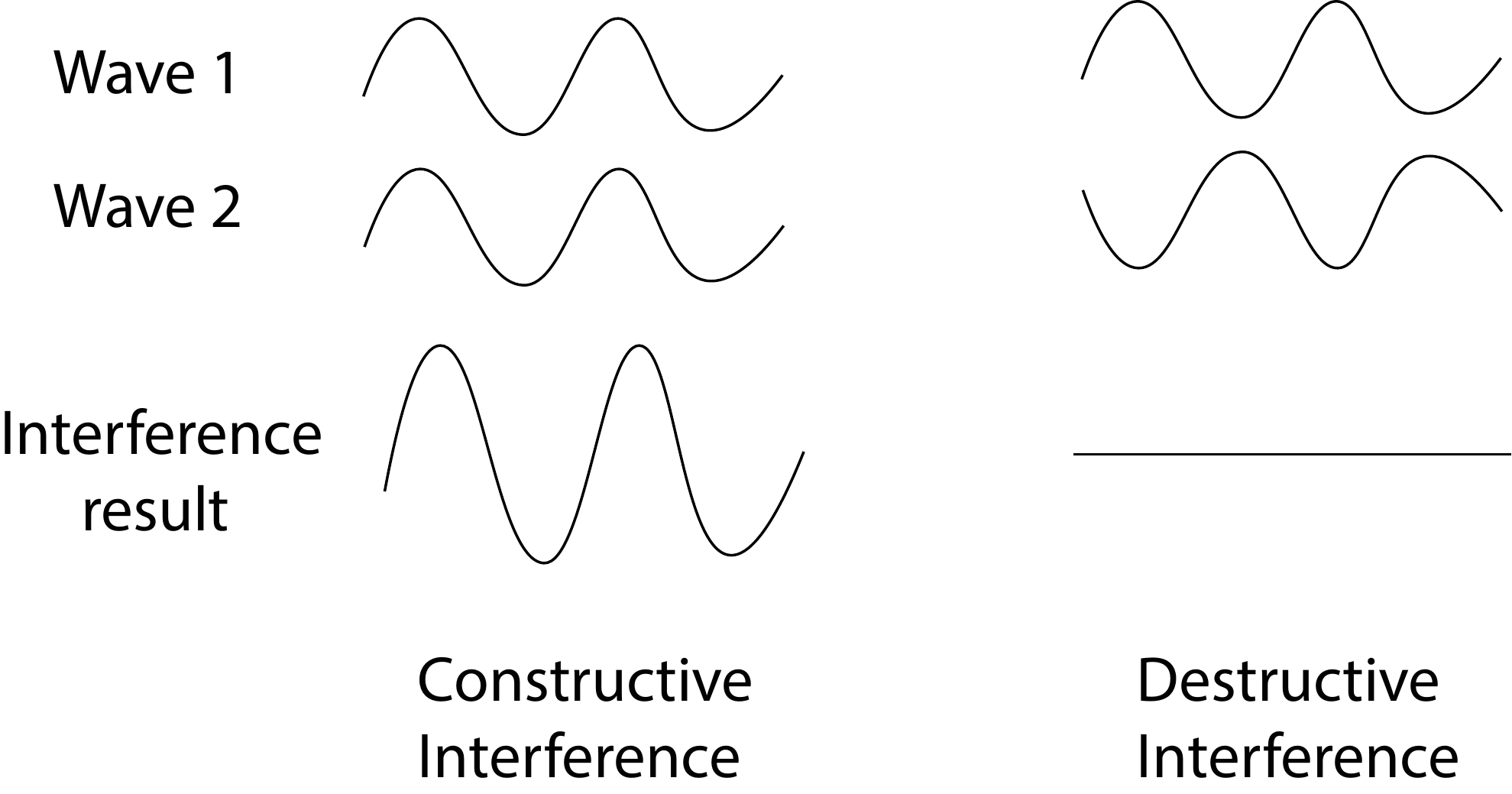}
  \caption{Wave Interference.}
  \label{fig:interference}
\end{figure}

\begin{figure}[t]
\centering
  \includegraphics[width=0.6\linewidth]{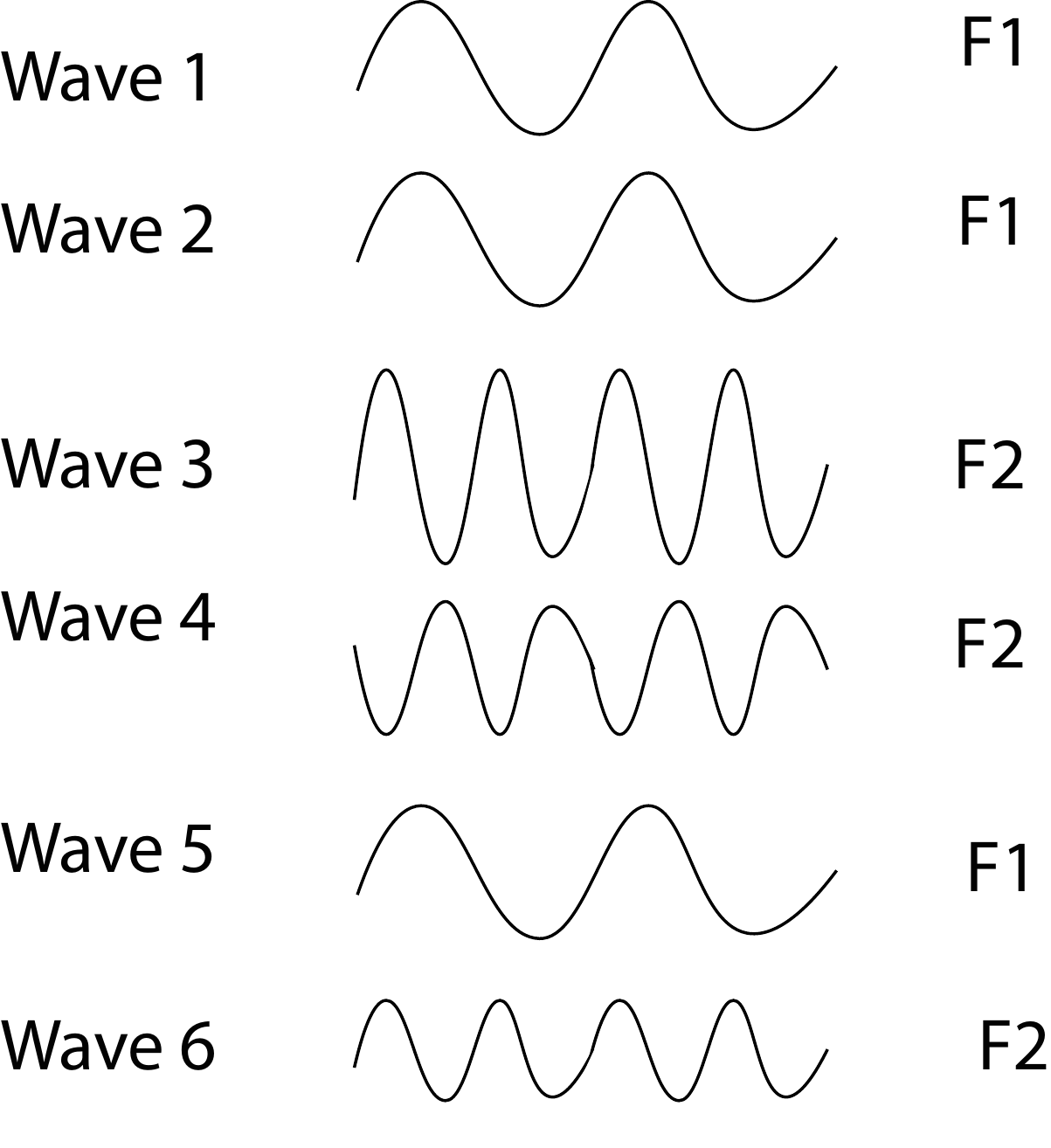}
  \caption{Different Frequency, Wavelength, and Amplitude Spin Wave Interference.}
  \label{fig:interferenceb}
\end{figure}

An excited SW is characterised by its wavelength $\lambda$ (the shortest distance between similar consecutive spins), wave number $k$ $\left(k=\frac{2*\pi}{\lambda}\right)$, frequency $f$ (determined by the complete spin precession time), phase $\phi$, and amplitude $A$, as  graphically indicated in Figure \ref{fig:SW_parameters}. As such, an SW can carry information encoded in its amplitude, phase, frequency, or a combination of them. Once formed, the SW propagates through the ferromagnetic material (waveguide) and may eventually meet other SWs present in the waveguide, case in which their interaction is governed by the wave interference principles. For instance, if two SWs with the same amplitude, wavelength, and frequency coexist in a waveguide, they interfere constructively if they have the same phase, and destructively if they are out of phase ($\Delta \phi=\pi$) as depicted in Figure \ref{fig:interference}. Furthermore, if more than two waves having the same $A$, $f$, and $\lambda$ interfere in the waveguide, the outcome captures a majority decision, i.e.,  if the number of spin waves having $\phi = 0$ is larger than the number of spin waves having $\phi = \pi$, the resulting spin wave has $\phi = 0$, and $\phi = \pi$ otherwise.  Thus, SW interference provides natural support for direct Majority gate implementations, e.g., $3$-input Majority is evaluated  by means of a $3$-SW interference in a waveguide \cite{logic1}, while its CMOS based implementation requires $18$ transistors. Moreover, SWs with different frequencies can coexist and propagate in the same waveguide without affecting each other and only interacting with other same-frequency SWs, which indicates that SW interaction provides intrinsic support for data parallel computing. Note that, in the most general case, spin waves with different amplitude, frequency, and wavelength can coexist and selectively interfere in the same waveguide, which results in more complex interference patterns as presented in Figure \ref{fig:interferenceb}. As depicted in the Figure, $f_1$ waves F1 and F2 interference results in F5 and  $f_2$ waves F3 and F4 interference results in F6, while no interaction between the $f_1$ and $f_2$ waves occurs. We note that in our investigation we consider that regardless of their frequency all input SWs have the same amplitude.

Depending on the orientation relation between spin wave propagation, effective magnetic field, and magnetization three main Magnetostatic Spin Wave (MSW) types exist: Magnetostatic Surface Spin Wave (MSSW), Forward Volume Magnetostatic Spin Wave (FVMSW), and Backward Volume Magnetostatic Spin Wave (BVMSW) \cite{Magnonic_crystals_for_data_processing}. While each type has certain interesting properties, FVMSWs are the most attractive as in-plane spin-wave propagation is isotropic, which is beneficial from the circuit design prospective. 

\begin{figure}[b]
\centering
  \includegraphics[width=0.4\linewidth]{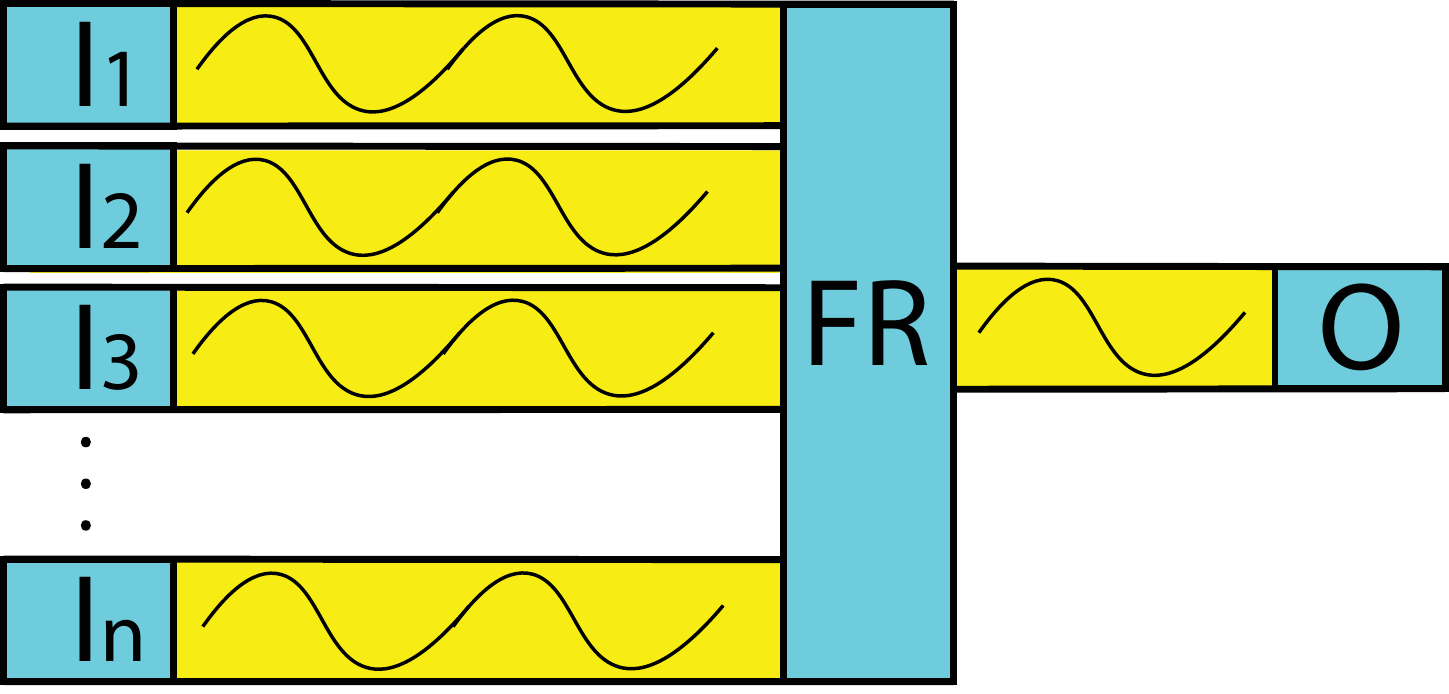}
  \caption{Conventional SW Logic Gate Structure}
  \label{fig:conventional_structure}
\end{figure} 

Figure \ref{fig:conventional_structure} depicts the generic structure of a SW based logic gate, which consists of multiple inputs ($I_1$, $I_2$, $I_3$, ..., $I_n$), a Functional Region (FR), which might perform Majority, AND, OR, XOR function or its inverted version, and an output $O$. All inputs are excited at the same frequency, propagate from their sources through the waveguide and interfere constructively or destructively based on their phases. The result is available at the output as a SW with the same frequency as the inputs. This is a scalar gate as each input SW represents one bit, thus in case the same function has to be pairwise evaluated on $n$-bit inputs this can be done in parallel by instantiating $n$ such gates or serially  by using one gate only with the associated area and delay overhead, respectively. In the following section we take advantage of different frequency SW interaction behaviour and introduce data parallel SW gates that can process $n$-bit inputs without hardware replication or serialisation.

\section{$n$-bit Data Parallel SW Logic Gate}
\label{sec:Proposed Parallelism Structure}
\begin{figure}[t]
\centering
  \includegraphics[width=0.6\linewidth]{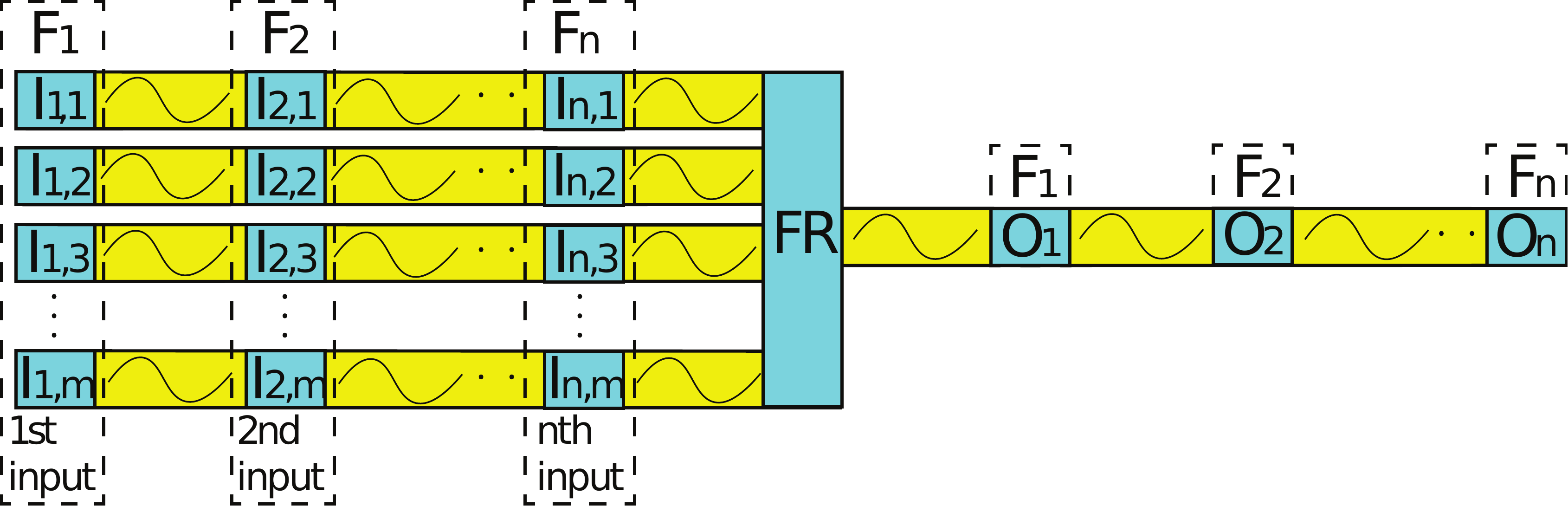}
  \caption{Multi-Frequency Spin Wave Logic Gate}
  \label{fig:proposed_architicture}
\end{figure}

Figure \ref{fig:proposed_architicture} presents the parallel spin wave logic gate we introduced in \cite{mahmoudDATE2020parallelism}, which is able to concurrently process $m$ $n$-bit inputs.  
As indicated in the Figure, the input sets ${\cal I}_i = \{I_{i,1}$, $I_{i,2}$, $I_{i,3}$, \ldots, $I_{i,m}\}, i = 1,2, \ldots, n$, are simultaneously encoded into SWs with frequency $f_i$  by means of, e.g., Magnetoelectric (ME) cells or antennas.  Subsequently, the SWs corresponding the sets ${\cal I}_i, i = 1,2, \ldots, n$ propagate through the waveguide without affecting each other until reaching the Functional Region (FR). Once the $m \times n$ spin waves arrive at  FR,  equal-frequency spin waves interfere constructively and destructively depending on their phases, producig $n$ output SWs ${\cal O}_i = {\cal F}({\cal I}_i),i = 1,2, \ldots, n$,  where ${\cal F}$ is the gate function, e.g., AND, OR, XOR. Those SWs can be sensed and transformed into the voltage domain by the detection cells located at $O_1$, $O_2$, \ldots, $O_n$ or tansmitted to the next SW gate. 

\begin{figure}[t]
\centering
  \includegraphics[width=0.6\linewidth]{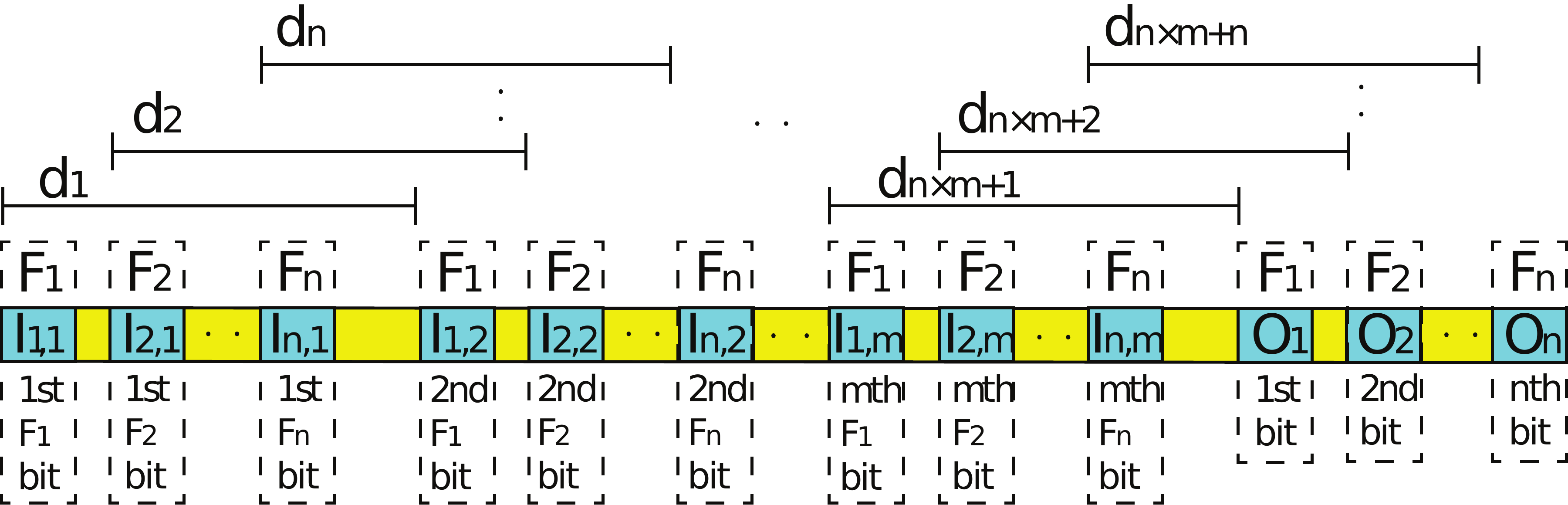}
  \caption{$n$-bit Inputs In-line Spin Wave Logic Gate}
  \label{fig:proposed_structure}
\end{figure}

Although the approach in Figure \ref{fig:proposed_architicture} is generic its practical realization requires stacked waveguides and contains bent regions, which impede smooth SW propagation.  We address these issues by  apply the same idea on a single waveguide structure and constructing the in-line gate in Figure \ref{fig:proposed_structure}. 

Note that for proper gate operation, SWs with the same frequency must be excited with the same amplitude and wavelength. Moreover, the distances between input sources and interference locations are SW frequency specific and crucial for proper gate functionality, thus they must be accurately determined. For example, if constructive interference is required for in-phase SWs and destructive for out of phase SWs, the distances between the same frequency sources must be $j _q \times \lambda_i, i=(1,2,3, \ldots, n)$, i.e, $d_1 = j_1\lambda_1$, $d_2=j_2 \lambda_2$, \ldots, $d_{nm}=j_{nm} \lambda_n$, where $j_q=\{1, 2, 3, \ldots\}, q=1,2,3, \ldots, nm$. Note that to minimize gate area and delay $j_q=1$ is the preferred choice, which is feasible for scalar gates but not always possible for parallel gates. 
Whereas, the distances must be $(j_q + \frac{1}{2}) \lambda_i$, i.e., $d_1=(j_1 + \frac{1}{2}) \lambda_1$, $d_2=(j_2 + \frac{1}{2}) \lambda_2$, \ldots, $d_{nm}=(j_{nm} + \frac{1}{2}) \lambda_n$, if the opposite behaviour is desired. 

In view of the previous discussion each output wave ${\cal O}_i$ is available for detection after a delay determined by the distance between the most faraway input cell of the ${\cal I}_i$ set, i.e., $I_{i,1}$ in Figure \ref{fig:proposed_structure}, and the output cell $O_i$, thus full parallelism is achieved. Note that the actual gate delay value can be optimized by choosing appropriate, e.g.,  waveguide material, dimensions, thickness, as discussed in Section \ref{sec:Simulation setup and experiments}. 

While delay optimization is a matter of  waveguide material and geometry choice, the gate area can be minimized by changing the  position of the input and output transducers. One can observe in Figure \ref{fig:proposed_structure} that input and output cells are ordered by bit position for clarity purpose. However, they can be shuffled as long as the previously discussed constraints are still satisfied, and this results in an area (overall gate length) reduction. 
To this end we introduce Algorithm 1, which identifies the transducer (source/detector) locations that are minimizing the waveguide length, while not infringing the wavelength dependent inter transducers distance constraints. The algorithm iteratively construct  the gate structure by instantiating one input set ${\cal I}_i, i =1, 2, \ldots, n$ at a time, while optimizing its transducer positions in relation to the already optimized structure embedding the previousely instantiated sets ${\cal I}_j, j = 1, 2, \ldots, i-1$.

The algorithm starts with a configuration in which all transducers are placed overlapped at the waveguide beginning. Subsequently, inputs sets are processed one at a time by initially placing them one after the other at $D$ distance regardless of the wavelength of the SW they process (line $3$ to $7$). If the first set was the one currently processed no further adjustments are required and the second set can be considered for placement.  If this is not the case, the for loop (line $9$ to $24$) is repositioning the transducer at the correct positions, which are multiples of their wavelength frequency. After this step, the transducer configuration for the up to date processed sets is the same as in Figure \ref{fig:proposed_structure}. Next, the for loop (line $25$ to $38$) performs the area optimization by checking the spaces between transducers and if it is possible moving one transducer if its wavelength imposed distance condition is satisfied. If one transducer has been moved {\bf Sort} reorders the transducers in the {\bf TP} matrix to capture the new configuration. These steps are repeated until all sets are placed and the gate length optimized. At the end, the gate area is calculated by multiplying the waveguide width by the waveguide length.

\begin{figure}
\centering
  \includegraphics[width=1.1\linewidth]{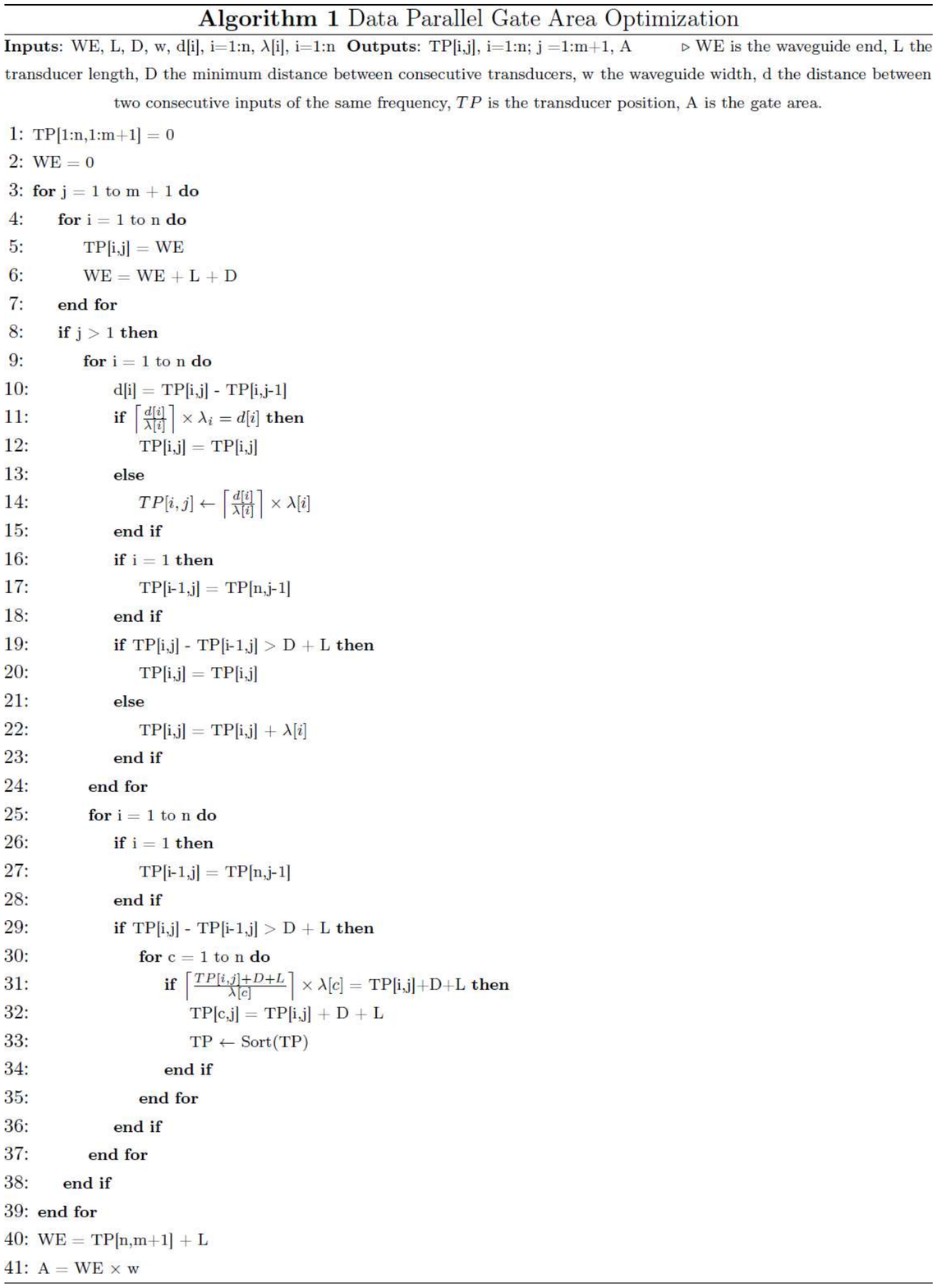}
\end{figure}

Let us assume a $3$-bit $2$-input gate operating on SWs with wavelength $\lambda_1$=\SI{100}{nm}, $\lambda_2$=\SI{50}{nm}, and $\lambda_3$=\SI{19}{nm},  \SI{10}{nm} transducer length, and \SI{1}{nm} minimum distance between transducers. By following the structure in Figure \ref{fig:proposed_structure}, the second input set can begin at \SI{33}{nm} 
from the waveguide start because the first three sources  $I_{1,1}, I_{1,2}, I_{1,3}$ occupy each \SI{10}{nm} and are \SI{1}{nm} distanced apart. As such the initial order is ($I_{1,1}, I_{1,2}, I_{1,3}, I_{2,1}, I_{2,2}, I_{2,3}, O_{1}, O_{2}, O_{3}$) with a corresponding  waveguide length of \SI{288}{nm}.  The optimization algorithm changes the order to ($I_{1,1},I_{1,2}, I_{1,3}, I_{2,3}, I_{2,2},I_{2,1}, O_{3},O_{2}, O_{1}$), which corresponds to  a \SI{210}{nm} waveguide length thus about $27$\% area savings.

Furthermore, two main methods can be utilized for output detection: (i) Phase detection, and (ii) Threshold detection. In the first case, a predefined phase is utilized as reference and a phase difference of $0$ represents a logic $0$, and a phase difference of $\pi$ a logic $1$. The second detection method assesses the SW magnetization (SWM) value and reports a $0$ logic if the SWM is smaller than a predefined threshold value and a logic $1$ otherwise.  If phase detection is in place, the gate can provide non-inverted or inverted output (or even both of them) by adjusting the reading location. 
For instance, referring to Figure \ref{fig:proposed_structure}, the detectors must be placed at a distance equal to (from the last $f_i$ SW source) $(j_q+\frac{1}{2}) \lambda_i, i=(1,2,3, \ldots, n)$,
such that $d_{nm+1}=(j_{nm+1}+\frac{1}{2}) \lambda_1$, $d_{nm+2}=(j_{nm+2}+\frac{1}{2}) \lambda_2$, \ldots, $d_{nm+n}=(j_{nm+n}+\frac{1}{2}) \lambda_n$, if the non-inverted results are desired. However, the detectors must be placed at a distance equal to (from the last $f_i$ SW sources)  $j \lambda_i$ such that $d_{nm+1}=j_{nm+1}\lambda_1$, $d_{nm+2}=j _{nm+2}\lambda_2$, \ldots, $d_{nm+n}=j_{nm+n} \lambda_n$ if the compliment is required. 
In the case of threshold based detection, the gate can provide non-inverted or inverted outputs without changing the output detector position by just switching the thresholding condition in the detector cell. Note that, regardless of the detection method, each read location should be as close as possible to the last input in its set to diminish the due to damping SW energy lost and process high amplitude spin waves.

\section{Simulation Setup}
\label{sec:Simulation setup and experiments}
This section provides inside on the simulation platform, parameters, and performed  experiments. 

\subsection{Simulation Platform}
The Object Oriented MicroMagnetic Framework (OOMMF) \cite{OOMMF} is utilized to evaluate the proposed structures. OOMMF numerically solves the LLG equation to capture the gate behaviour. The OOMMF input  is a TCL/TKL script, which describes the gate and the input stimuli and the results can be visualized within the OOMMF framework or post-processed by other tools like matlab, which is the case in this paper. 

\subsection{Simulation Parameters}

$Fe_{60}Co_{20}B_{20}$ waveguides that have waveguide width of \SI{50}{nm} with Perpendicular Magnetic Anisotropy (PMA) are utilized for all gate constructions. We note that for this material the anisotropy field $H_{anisotropy} > M_s$, which means that there is no need for the application of an external magnetic field \cite{parameters}. Table \ref{table:1} presents the parameter we utilize to validate the $8$-bit $2$-input XOR/XNOR and $3$-input Majority gates. The $8$ SW frequencies are  \SI{10}{GHz}, \SI{20}{GHz},  \SI{30}{GHz}, \SI{40}{GHz}, \SI{50}{GHz}, \SI{60}{GHz},  \SI{70}{GHz}, and \SI{80}{GHz}. By making use of the FVMSW dispersion relation and given that the wavenumber $k=\frac{2\pi}{\lambda}$, we determine the distances between transducers exciting/detecting SWs with the same frequency are: $d_1$=\SI{166}{nm} (j=2), $d_2$=\SI{100}{nm} (j=2), $d_3$=\SI{117}{nm} (j=3), $d_4$=\SI{165}{nm} (j=5), $d_5$=\SI{174}{nm} (j=6), $d_6$=\SI{130}{nm} (j=5), $d_7$=\SI{168}{nm} (j=7), and $d_8$=\SI{176}{nm} (j=8). Furthermore, an \SI{1}{nm} minimum separation distance between transducers is in place. 

\begin{table}
\caption{Parameters}
\label{table:1}
\centering
  \begin{tabular}{|c|c|}
    \hline
    Parameters & Values \tabularnewline
    \hline
    Magnetic saturation $M_s$ & $1.1$ $\times$ $10^6$ \SI{}{A/m} \tabularnewline
    \hline
    Perpendicular anisotropy constant $k_{ani}$ & \centering $8.3177$ $\times$ $10^5$ \SI{}{J/m^3} \tabularnewline
    \hline
    Damping constant $\alpha$ & $0.004$ \tabularnewline
    \hline
    Waveguide thickness $t$ & \SI{1}{nm} \tabularnewline
    \hline
    Exchange stiffness $A_{exch}$ & \SI{18.5}{pJ/m} \tabularnewline
    \hline
  \end{tabular}
\end{table}

\subsection{Performed Simulations}

We perform the following simulation experiments: 

\begin{itemize}
\item $8$-bit $2$-input XOR/XNOR gate with threshold detection. The two $8$-bit inputs are simultaneously excited using the sources ($I_{1,1}, I_{2,1}, I_{3,1}, \ldots, I_{8,2}$). The excited spin waves propagate through the waveguide and those who have the same frequencies interfere with each other. The resulting spin waves propagate towards the output where they are captured at $O_1, O_2, \ldots, O_8$ based on threshold detection. We carry on the validation of both area unoptimized ($I_{1,1}, I_{2,1}, I_{3,1}, I_{4,1}, I_{5,1}, I_{6,1}, I_{7,1}, I_{8,1}$, $I_{1,2}, I_{2,2}, I_{3,2}, I_{4,2}, I_{5,2}, I_{6,2}, I_{7,2}, I_{8,2}, I_{1,3}, I_{2,3}, I_{3,3}, I_{4,3}$, $I_{5,3}, I_{6,3}, I_{7,3}, I_{8,3}$) and optimized ($I_{1,1}, I_{2,1}, I_{3,1}, I_{4,1}, $ $I_{5,1}, I_{6,1}, I_{7,1}, I_{8,1}, I_{2,2}, I_{3,2}, I_{1,2}, I_{6,2}, I_{4,2}, I_{5,2}, I_{7,2}, I_{8,2}, $ $I_{2,3}, I_{8,3}, I_{3,3}, I_{1,3}, \\ I_{6,3}, I_{4,3}, I_{5,3}, I_{7,3}$) configurations.  Note that as  detectors order is not important they follow the same pattern, i.e., ($O_{1}, O_{2}, O_{3}, O_{4}, O_{5}, O_{6}, O_{7}, O_{8}$) in both cases. 
\item $8$-bit $3$-input {Majority} gate based on phase detection. We again considered area unoptimized and optimized gate instances but in this case detector order is relevant, thus the after optimization source and detector order is $I_{1,1}, I_{2,1}, I_{3,1}, I_{4,1}, I_{5,1}, I_{6,1}, I_{7,1}, I_{8,1}, I_{2,2},\\ I_{3,2},  I_{1,2}, I_{6,2}$, $I_{4,2}, I_{5,2}, I_{7,2}, I_{8,2}, I_{2,3}, I_{8,3}, I_{3,3}, I_{1,3}, I_{6,3}, I_{4,3}, I_{5,3}, I_{7,3}$, $O_{6}, O_{8}, O_{4}, O_{2}, O_{5}, O_{1},\\ O_{7}, O_{3}$.
\end{itemize}

\section{Simulation Results and Discussion}
\label{sec:Simulation results and discussion}
This section presents simulation results for the $8$-bit $2$-input XOR/XNOR and $3$-input Majority gate instances, performance estimations, and a comparison with SW state-of-the-art functionally equivalent structures. Subsequently,  it discusses fan-in and geometric scalability, and maximum achievable parallelism (upper bound of the number of practically achievable SW frequencies)  issues, and variability and thermal noise effects. 

\begin{figure}[t]
\centering
  \includegraphics[width=0.6\linewidth]{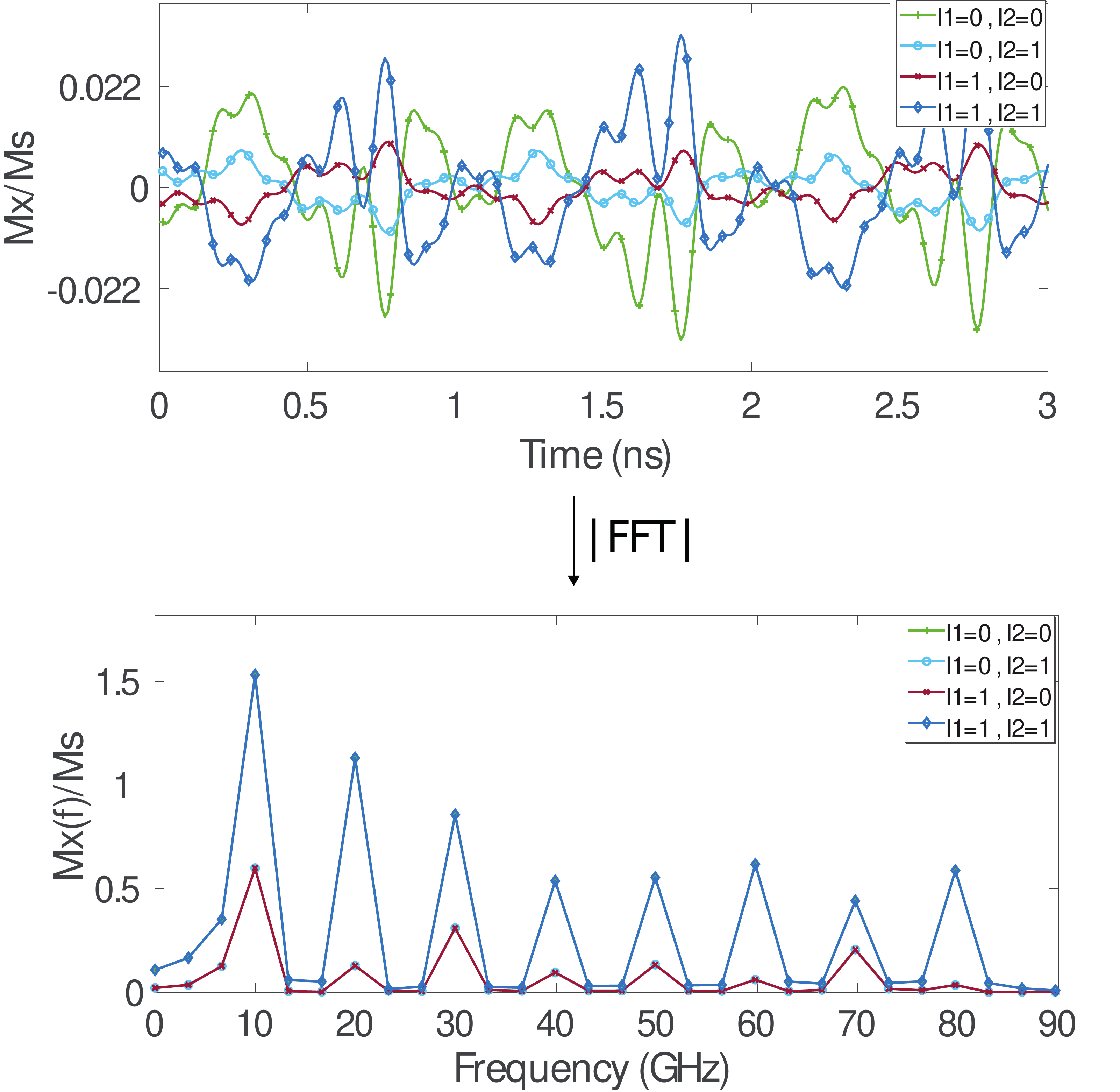}
  \caption{ Unoptimized $8$-bit XOR Gate Time and Frequency Response.}
  \label{fig:results1}
\end{figure}

\begin{figure}[t]
\centering
  \includegraphics[width=0.6\linewidth]{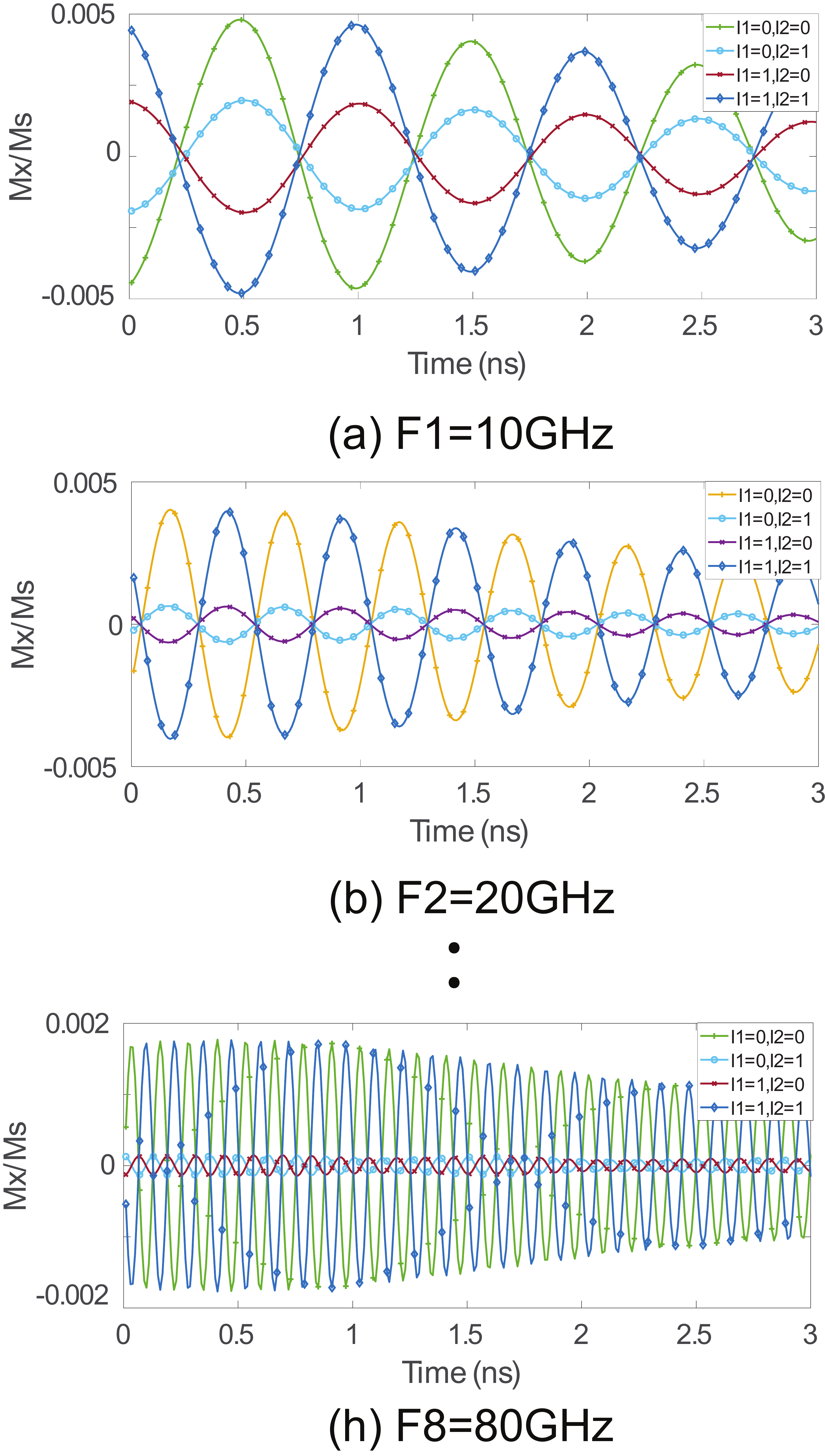}
  \caption{Unoptimized $8$-bit XOR Gate Outputs a) $f_1$=\SI{10}{GHz}, b) $f_2$=\SI{20}{GHz}, \ldots, h) $f_8$=\SI{80}{GHz}.}
  \label{fig:results2}
\end{figure} 

\begin{figure}[t]
\centering
  \includegraphics[width=0.6\linewidth]{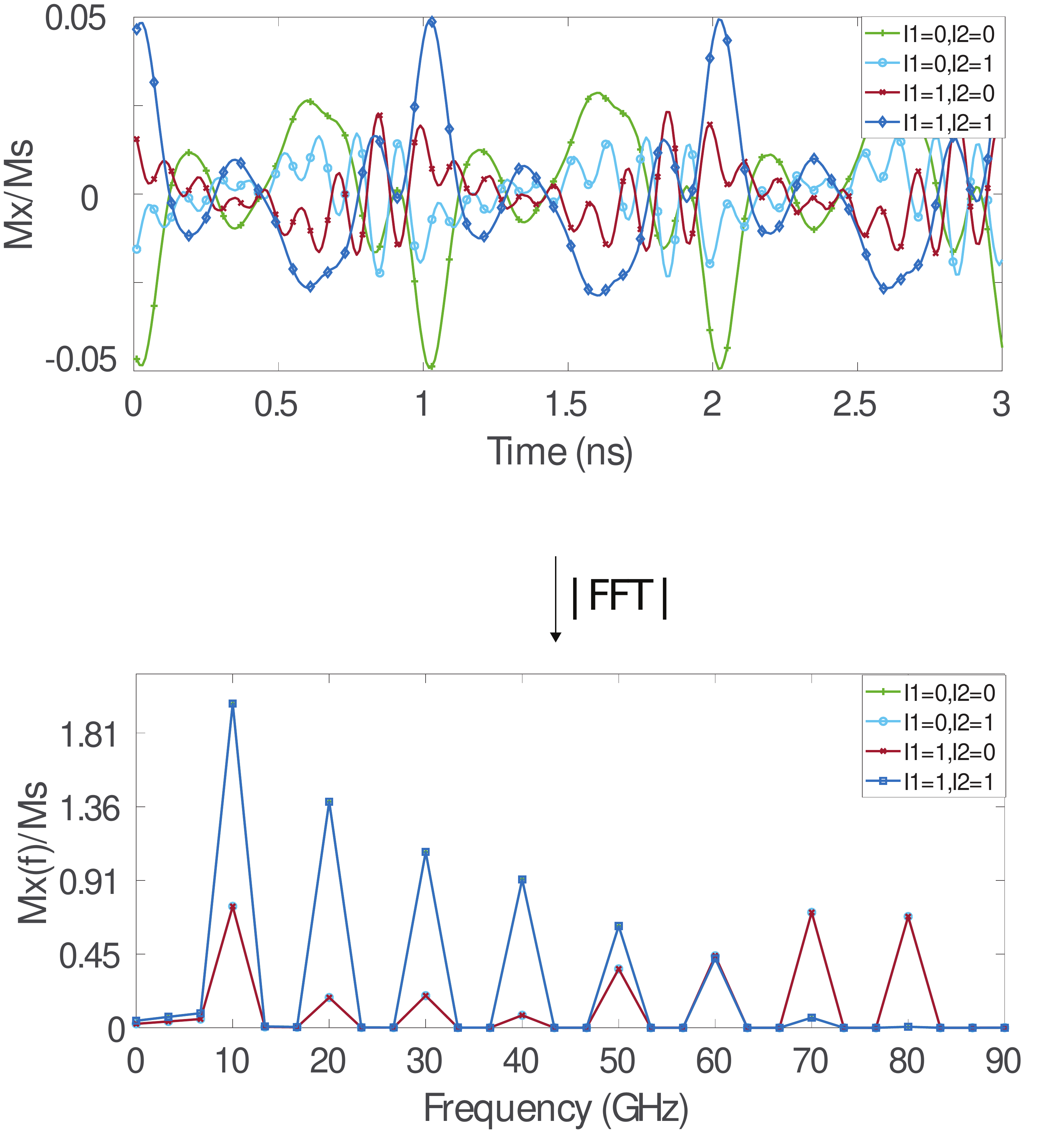}
  \caption{Optimized $8$-bit XOR Gate Time and Frequency Response.}
  \label{fig:results3}
\end{figure} 

\begin{figure}[t]
\centering
  \includegraphics[width=0.6\linewidth]{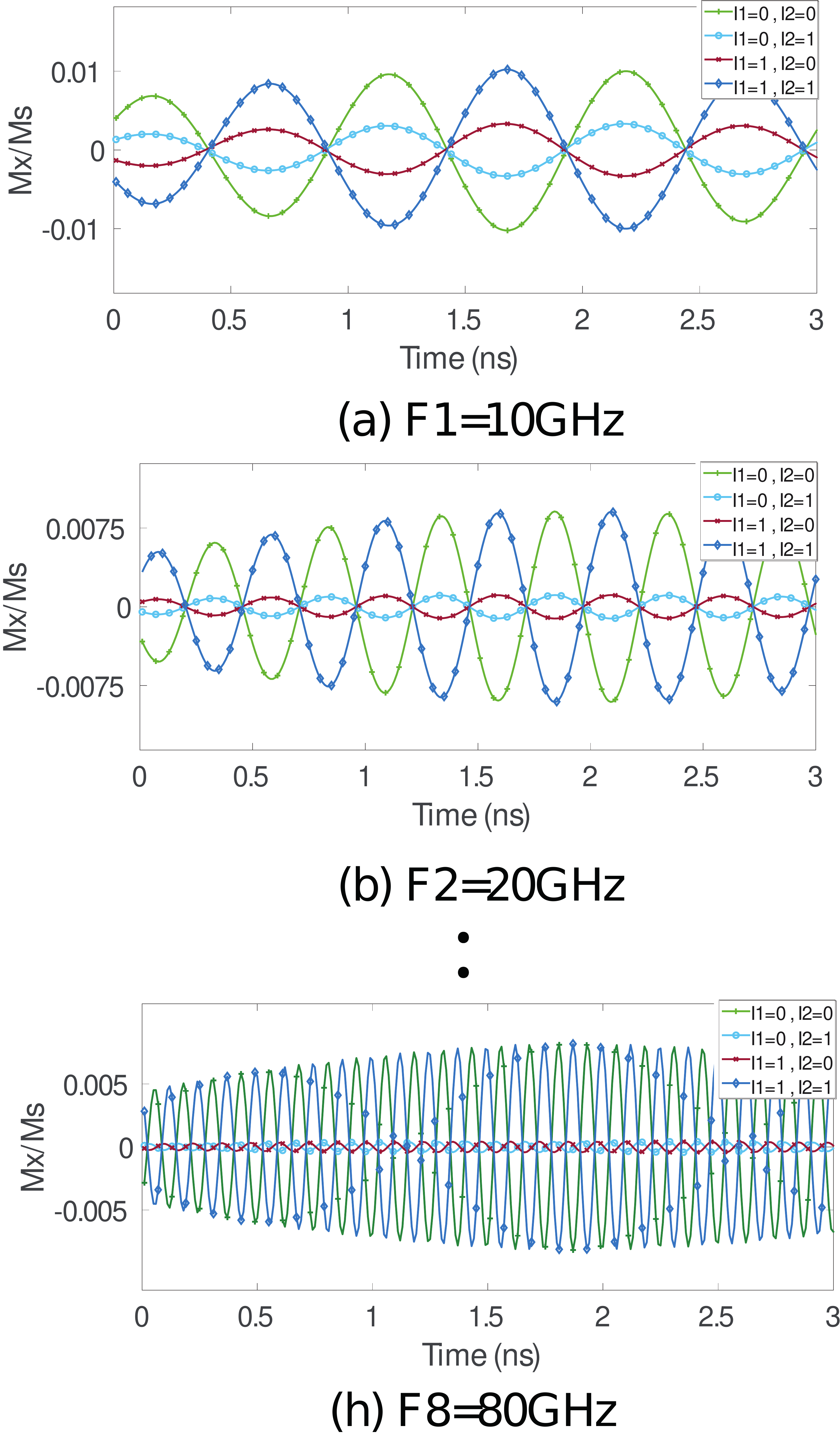}
  \caption{Optimized $8$-bit XOR Gate Outputs: a) $f_1$=\SI{10}{GHz}, b) $f_2$=\SI{20}{GHz}, \ldots, h) $f_8$=\SI{80}{GHz}.}
  \label{fig:results4}
\end{figure} 

\begin{figure}[t]
\centering
  \includegraphics[width=0.6\linewidth]{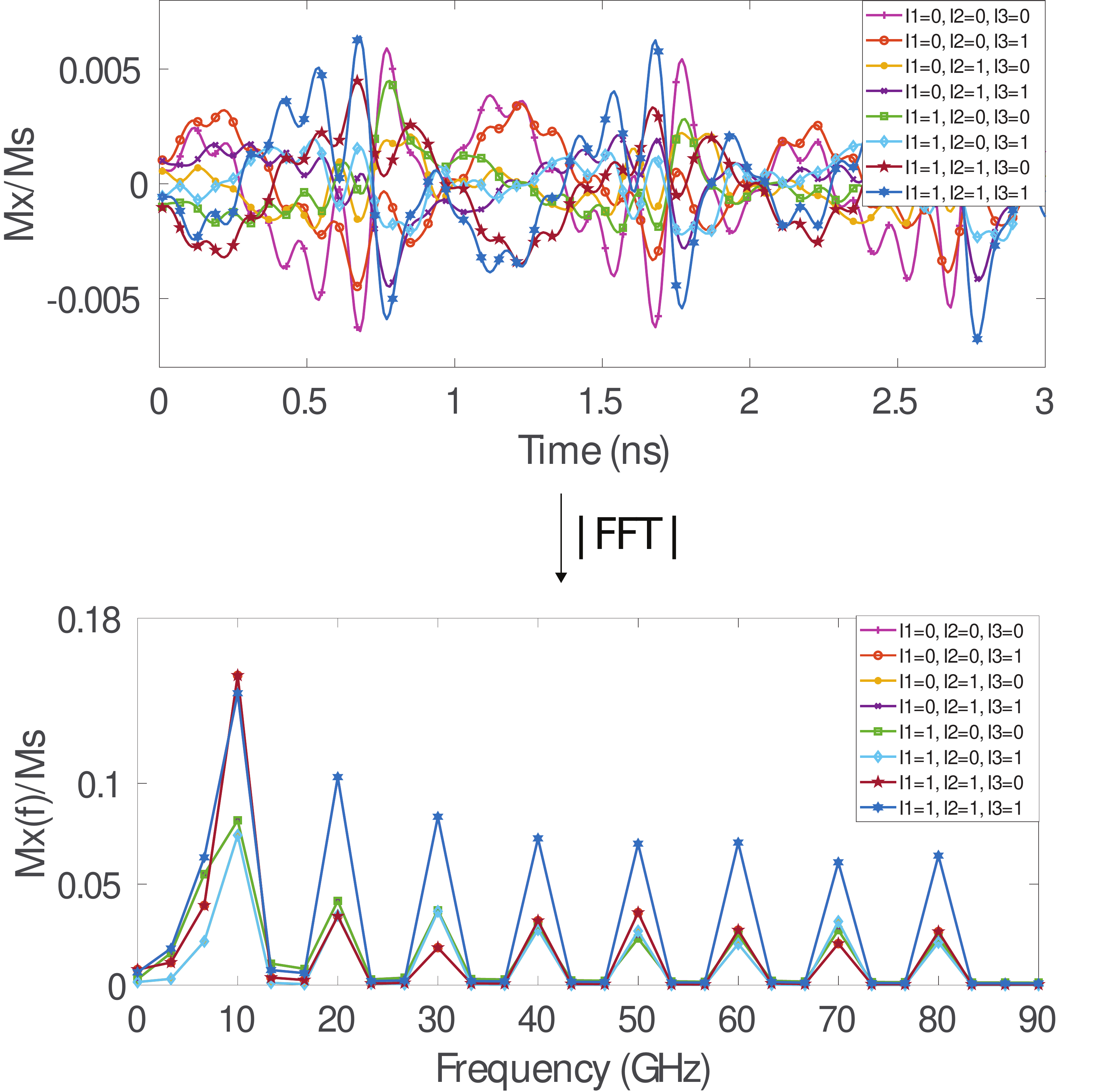}
  \caption{Unoptimized $8$-bit Majority Gate Time and Frequency Response.}
  \label{fig:results5}
\end{figure} 

\begin{figure}[t]
\centering
  \includegraphics[width=0.6\linewidth]{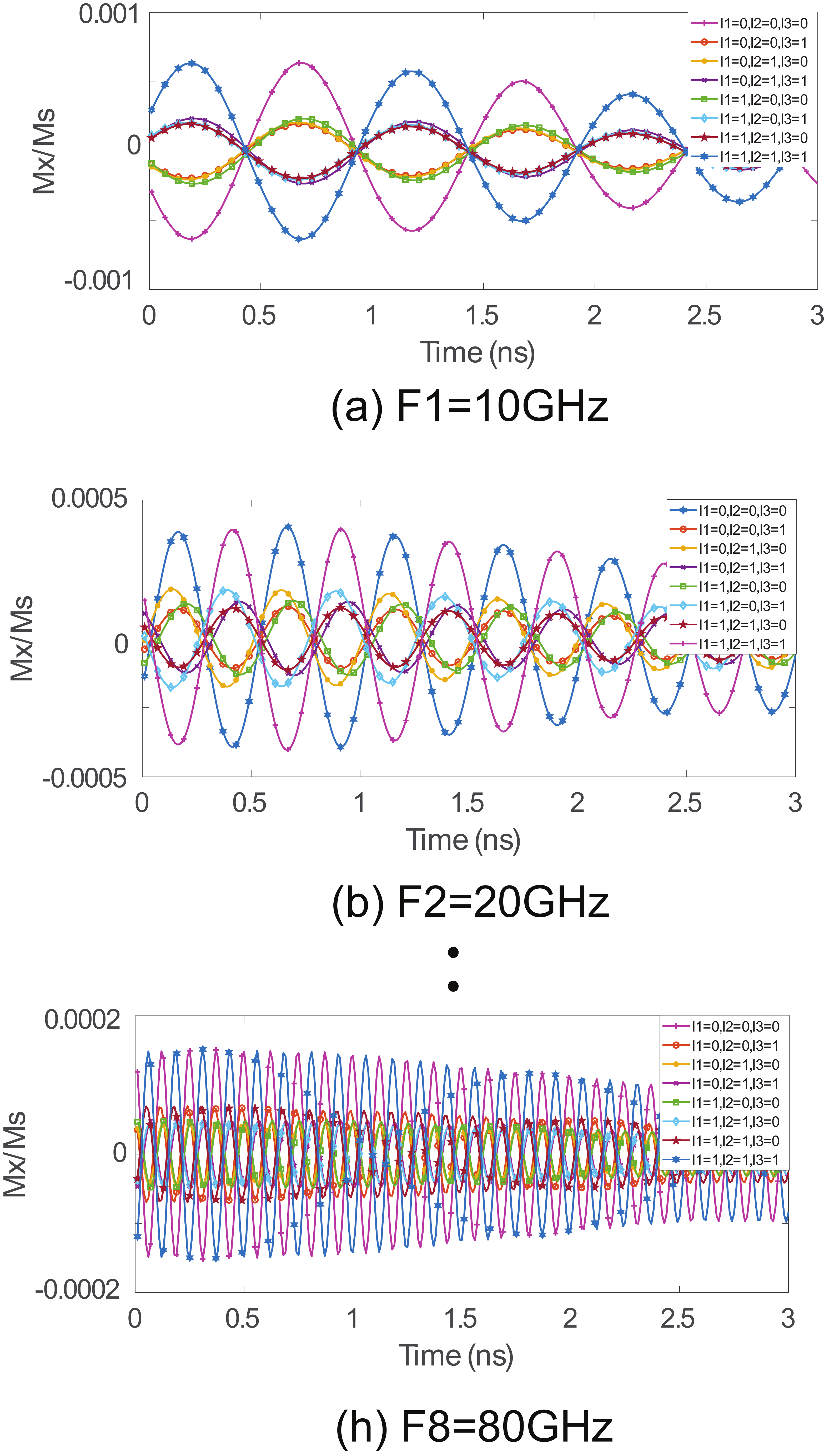}
  \caption{Unoptimized $8$-bit Majority Gate Outputs a) $f_1$=\SI{10}{GHz}, b) $f_2$=\SI{20}{GHz}, \ldots, h) $f_8$=\SI{80}{GHz}.}
  \label{fig:results6}
\end{figure} 

\begin{figure}[t]
\centering
  \includegraphics[width=0.6\linewidth]{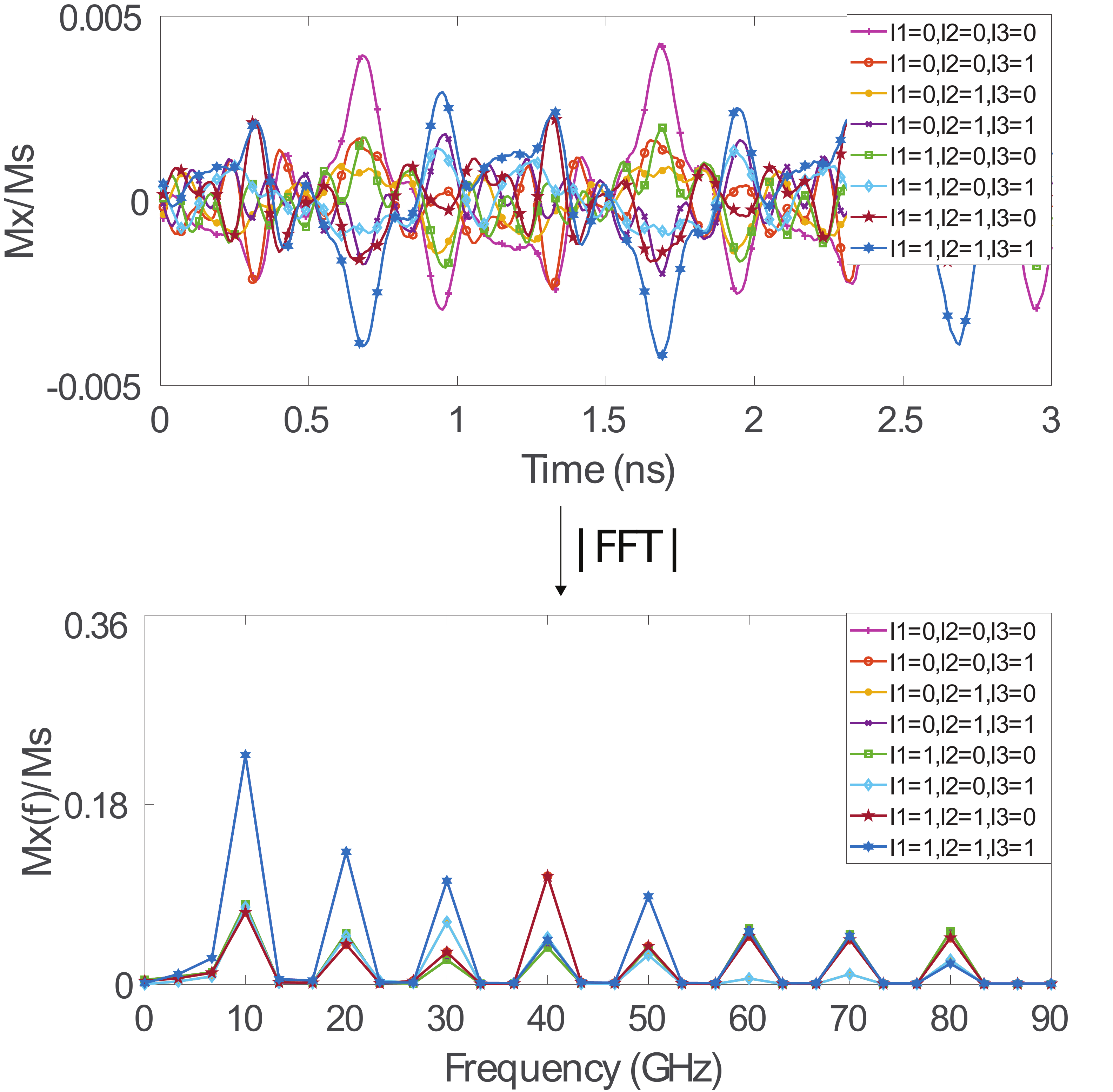}
  \caption{Optimized $8$-bit Majority Gate Time and Frequency Response.}
  \label{fig:results7}
\end{figure} 

\begin{figure}[t]
\centering
  \includegraphics[width=0.6\linewidth]{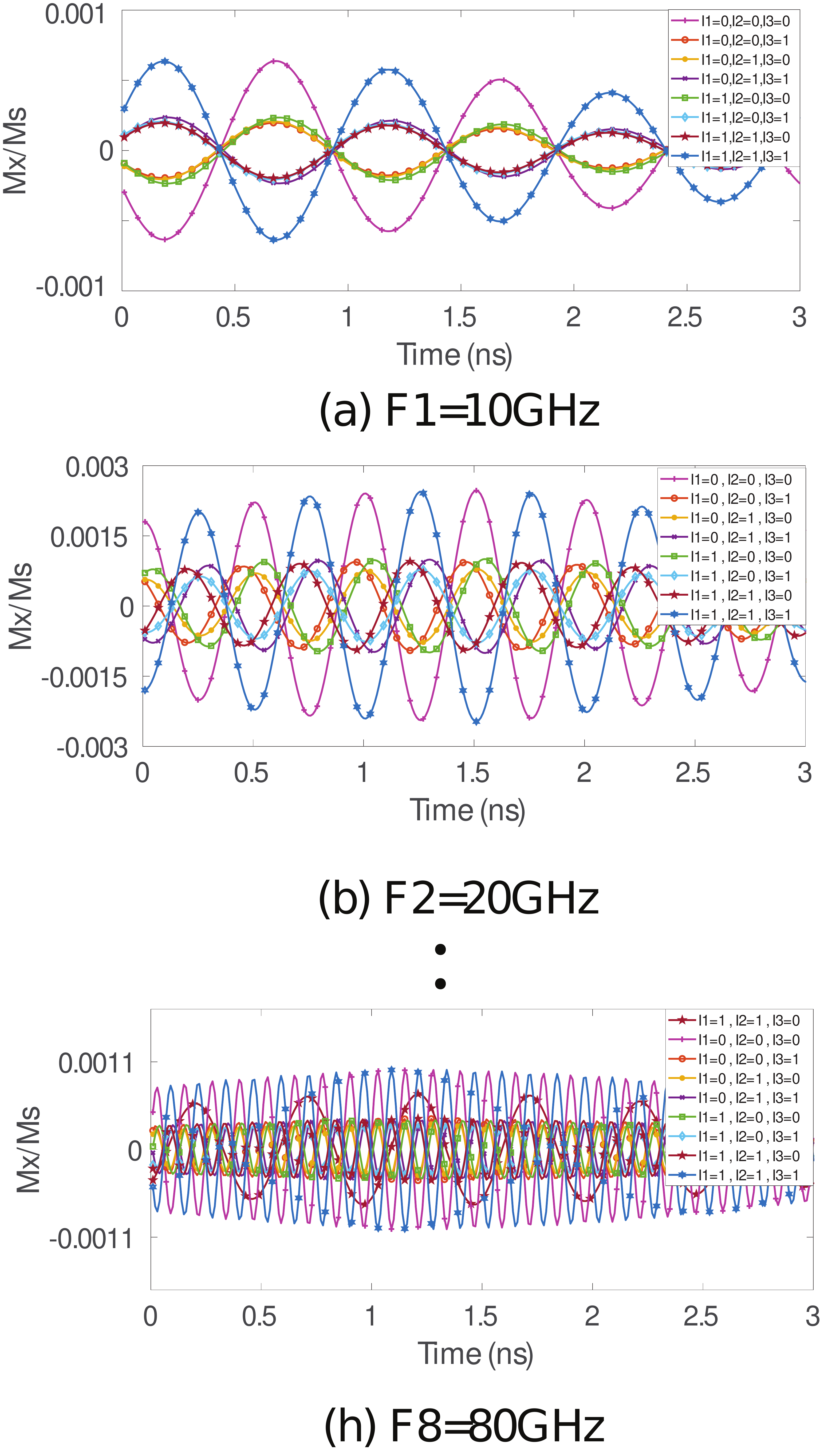}
  \caption{Optimized $8$-bit Majority Gate Outputs a) $f_1$=\SI{10}{GHz}, b) $f_2$=\SI{20}{GHz}, \ldots, h) $f_8$=\SI{80}{GHz}.}
  \label{fig:results8}
\end{figure} 

\subsection{Simulation Results}

\subsubsection*{$8$-bit $2$-input threshold detection based {XOR/XNOR} gate} Figure \ref{fig:results1} presents OOMMF simulation results for the area unotimized byte-based $2$-input XOR gate instance. The $y$-axis reflects the output SWs Mx over Ms ratio, i.e., magnetization in the $x$-direction over magnetic saturation.  To simplify the Figure we only assume all $0$s and all $1$s input sets, thus only four input combinations are possible, and as such the gate response to any input combination is the same in all frequencies. As expected same-frequency SW pairs interfere without affecting the other SWs and this is clear from Figure \ref{fig:results1}, which indicates that  $8$ different frequencies components exist without distorting each-other in the Fast Fourier Transform (FFT) amplitude spectrum for all the considered input combinations. Moreover, as it can be noticed from Figure \ref{fig:results2}, the output SWs are not distorted and can be properly detected for each frequency. Let us consider the first output detection cell, which is tuned for the  \SI{10}{GHz} SW. When reading the output at time \SI{0.5}{ns} for ${\cal I}_1={\cal I}_2=0$ and ${\cal I}_1={\cal I}_2=1$, the absolute SW magnetization value is greater than $0.0035$ $M_s$ due to the constructive interference, whereas the SW magnetization is less than $0.0035$ $M_s$ when one input set is $0$ and the other one is $1$. Therefore, if the detection threshold is set to $0.0035$ $M_s$ an XOR function is obtained as a SW magnetization greater (lower) than $0.0035$ $M_s$ is read as a logic $0$ ($1$). An XNOR can be realized by flipping the condition such that a SW magnetization lower (greater) than $0.0035$ $M_s$ is read as a logic $0$ ($1$). 
Similarly, for the second detection cell, which targets the  \SI{20}{GHz} SW a threshold value  of $0.0032$ $M_s$ is in place and by following a similar way of reasoning threshold values of $0.0028$ $M_s$, $0.0025$ $M_s$, $0.0022$ $M_s$, $0.0017$ $M_s$, $0.0015$ $M_s$, and  $0.001$ $M_s$ can be determined  for the rest of frequencies. 

Figure \ref{fig:results3} and \ref{fig:results4} present OOMMF simulation results for the optimized $8$-bit $2$-input XOR gate. As depicted in Figure \ref{fig:results4}, the simulation proves the correct functionality of the XOR/XNOR gate. One can observe in the Figure that in this case the SW magnetization at all frequencies is higher as the spin waves propagate on lower distances when compared with the non-optimized case. In addition, the detection threshold values are higher, i.e., $0.007$ $M_s$, $0.005$ $M_s$, $0.0045$ $M_s$, $0.0038$ $M_s$, $0.0034$ $M_s$, $0.0027$, $0.0025$ $M_s$, and $0.002$ $M_s$,  
therefore, less sensitive detectors are requited for the XOR/XNOR gate implementation.

\subsubsection*{$8$-bit phase detection based $3$-input {Majority} gate} The $8$-bit $3$-input unoptimized Majority gate OOMMF simulation results are presented in Figure \ref{fig:results5}. The same notations are in place and again, to simplify the Figure we only assume all $0$s and all $1$s input sets, thus only $8$ input combinations are presented.  
The Figure clearly demonstrate proper gate functionality as $8$ different frequencies components exist without distorting each-other in the Fast Fourier Transform (FFT) amplitude spectrum for all the possible input combinations (${\cal I}_1={\cal I}_2={\cal I}_3=0$), (${\cal I}_1={\cal I}_2=0, {\cal I}_3=1$), \ldots, (${\cal I}_1={\cal I}_2={\cal I}_3=1$). 
Figure \ref{fig:results6} indicates that the output SWs are not distorted and can be properly detected for each frequency. 
Let us concentrate on Figure \ref{fig:results6}a, which captures the \SI{10}{GHz} $3$-input Majority gate response and consider the output at time moment \SI{0.75}{ns}, When the three inputs have the same phase of $0$ ($I_1I_2I_3=000$) they constructively interfere in the waveguide resulting in a phase of $0$ SW, which corresponds to a logic $0$. Also, when at most one of the inputs is logic $1$ ($I_1I_2I_3=001$, $I_1I_2I_3=010$, $I_1I_2I_3=100$), i.e., has phase of $\pi$, the SWs interfere constructively and destructively, and the results is still a logic $0$. In contrast, if at most one of the inputs is logic $0$ ($I_1I_2I_3=011$, $I_1I_2I_3=110$, $I_1I_2I_3=101$), then the output is logic $1$ as a result of the interferences. Further, when the three inputs have the same phase of $\pi$ ($I_1I_2I_3=111$), then spin waves interfere constructively in the waveguide, which results in a phase of $\pi$, which corresponds to a logic $1$. The same line of reasoning can be applied for all the other $7$ cases as it is clearly indicated by Figure \ref{fig:results6}.

The optimized $8$-input $3$-input Majority gate OOMMF simulation results are presented in Figure \ref{fig:results7} and \ref{fig:results8}. As it can be observed from Figure \ref{fig:results8}, the gate functions correctly while the SW amplitudes are higher as due to the optimization SWs propagate over shorter distances, which enables the utilization of less sensitive detectors.

\subsection{Performance Evaluation}

To get inside on the practical potential of our proposal, we evaluate and compare the $8$-bit gates with functionally equivalent state-of-the-art SW implementation obtained by the instantiation of $8$ normal (scalar) Majority/XOR gates, in terms of area, delay, and power consumption. In our evaluations we make the following assumptions: (i) source/detector dimensions are \SI{10}{nm} $\times$ \SI{50}{nm} as suggested in \cite{mahmoudDATE2020parallelism}, (ii) SW propagation through the waveguide doesn't consume noticeable energy, and (iii) transducer delay is \SI{0.42}{ns} \cite{excitation_table_ref16}.

Under this assumptions we first evaluate the optimization algorithm impact on the $8$-bit gates area. Our calculations indicate that the unoptimized XOR and Majority gates have an area of \SI{0.02525}{\mu m^2} and \SI{0.04725}{\mu m^2}, respectively, which become \SI{0.01755}{\mu m^2} and \SI{0.0279}{\mu m^2}, respectively, after the optimization. This clearly proves  the algorithm efficiency as it diminishes the area by $30$\% and $41$\%, respectively.

As the standard functionally equivalent implementations require $8$ $2$-input XOR and $8$ $3$-input Majority gates it occupies  \SI{0.0784}{\mu m^2} and \SI{0.116}{\mu m^2} real estate, respectively, our proposal enables a $4.47$x and $4.16$x area reduction, respectively. 

Generally speaking, to calculate an SW gate delay one needs to sum-up the time associated to SW generation, propagation, and detection.  The due to SW propagation through the waveguide delay depends on the travelled distance from generation to detection and it can be computed by dividing the distance by the SW group velocity, which is \SI{3500}{m/s}  for CoFeB \cite{Magnonic_crystals_for_data_processing}. Given that the longest propagation path for the $8$-bit $2$-input XOR and $3$-input Majority gates is \SI{351}{nm} and \SI{558}{nm}, respectively, the propagation delay is \SI{100}{ps} and \SI{159}{ps}, respectively, which by adding the transducers delay sums up to  \SI{940}{ps} and \SI{999}{ps}, respectively. For the scalar $2$-input XOR and $3$-input Majority gates the longest path is \SI{196}{nm} and \SI{290}{nm}, respectively, which translates into a transmission delay of \SI{56}{ps} and \SI{83}{ps}, respectively, and  \SI{896}{ps} and \SI{923}{ps} overall gate delay, respectively. Thus, the $8$-bit $2$-input XOR and $3$-input Majority gates are slower than their scalar counterparts with $5$\% and $7$\%, respectively. 

As both parallel and scalar gate implementations make use of the same number of transducers and the through the waveguide  propagation consumes insignificant power, the two implementations are equivalent in terms of power consumption.

\subsection{Fan-in and Geometrical Scalability} 

The proposed structure is generic and the number of bits per frequency, i.e., the gate fan-in, shouldn't affect its functionality. However, as the number of inputs increases, the damping effect might play a more significant role in diminishing SW amplitudes. Thus, if a large number of inputs is targeted, it might be needed to excite the same frequency SW inputs in Figure \ref{fig:proposed_structure} at different energy levels $E_n$ $<$ $E_{n-1}$ $<$ \ldots $<$ $E_1$, where $E_i$ is the energy that the $i^{th}$ SW is excited at. We note however that: (i) usual fan-in values are rather small ($2$ and $3$ in the gates we designed), (ii) energy level differentiation is only required for large fan-in values in case the logic gate doesn't function correctly, and (iii) within certain limits the SW energy levels can be adjusted by properly biasing the source transducers. 

To get inside on the effect of the waveguide width on gate functionality we scaled it from \SI{50}{nm} up to \SI{500}{nm} 
It was noticed that scaling doesn't affect the gates functionality and it doesn't generate any crosstalk effects. We note that, as waveguide width increases, the ferromagnetic resonance frequency decreases and thus lower SW frequencies can be utilized. Although this is advantageous from signal loss perspective such structures require  stronger static magnetic fields, which results in area and energy consumption overheads.

\begin{figure}[t]
\centering
  \includegraphics[width=0.8\linewidth]{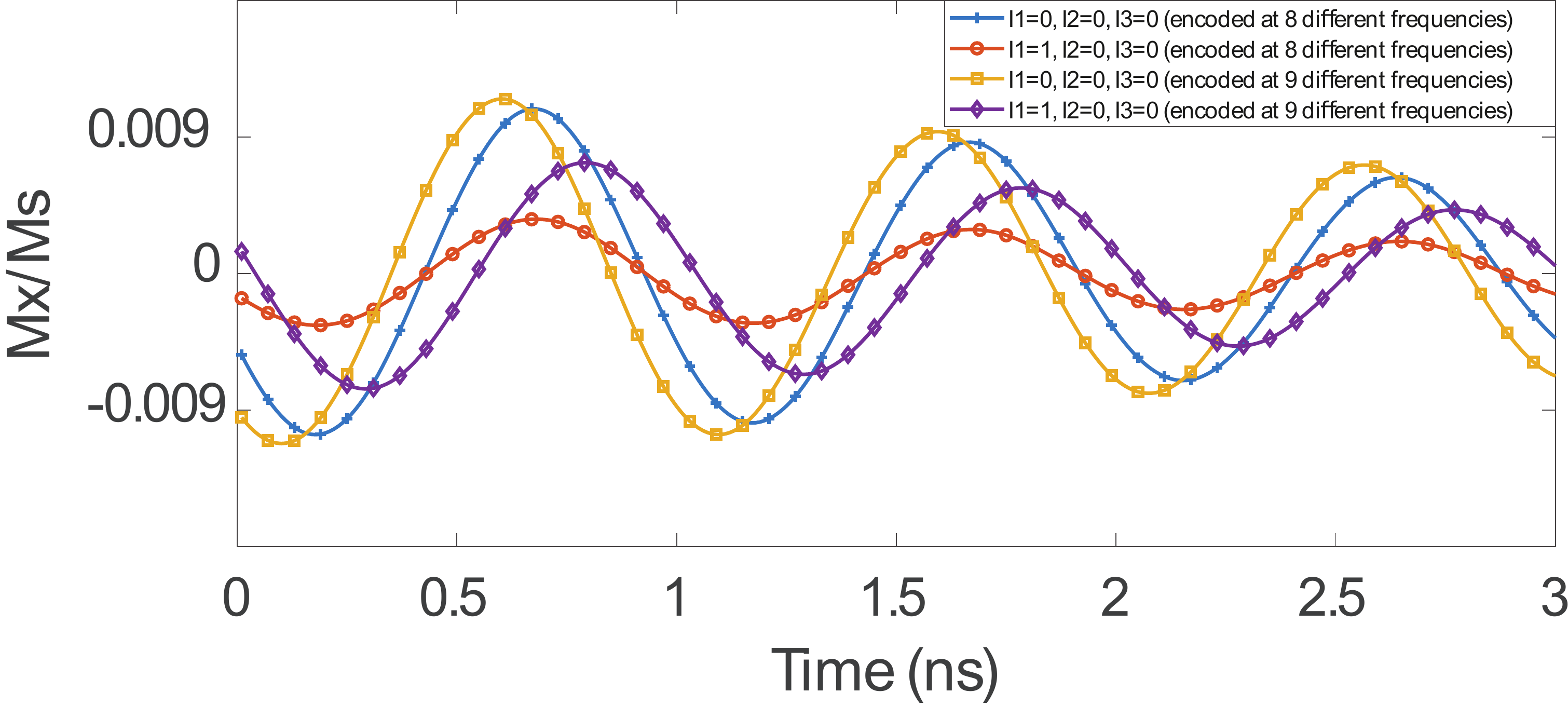}
  \caption{MAJ Gate Outputs at $f_1$=$10$GHz.}
  \label{fig:results9}
\end{figure} 

\subsection{Practically Achievable Parallelism} 

To get some inside on the data parallelism practical upper-bound we examined the consequences of increasing the number of bits per set, i.e., utilized frequencies. To this end we OOMMF simulate $8$-bit and $9$-bit $3$-input Majority gate instances and  display in Figure \ref{fig:results9} the $10$ GHz frequency output component for the input combinations ${\cal I}_1{\cal I}_2 {\cal I}_3=000$ and ${\cal I}_1{\cal I}_2 {\cal I}_3=100$ . 

One can observe in the Figure that at time=\SI{0.5}{ns} the $8$-bit Majority gate output has the same phase for the considered input combination, which reflects the correct functionality of the Majority gate as in both cases $0$ is the majority. 
However, the  $9$-bit Majority gate output at time=\SI{0.5}{ns} has different phase, $0$ for ${\cal I}_1{\cal I}_2 {\cal I}_3=000$, and approximately $\pi/4$ for ${\cal I}_1{\cal I}_2 {\cal I}_3=100$, which indicate that the gate starts to malfunction.  Based on this we can conclude that, for the proposed topology and utilized material, $8$ is the maximum number of frequencies one can use to construct robust parallel SW gates. 

\begin{figure}
\centering
  \includegraphics[width=0.8\linewidth]{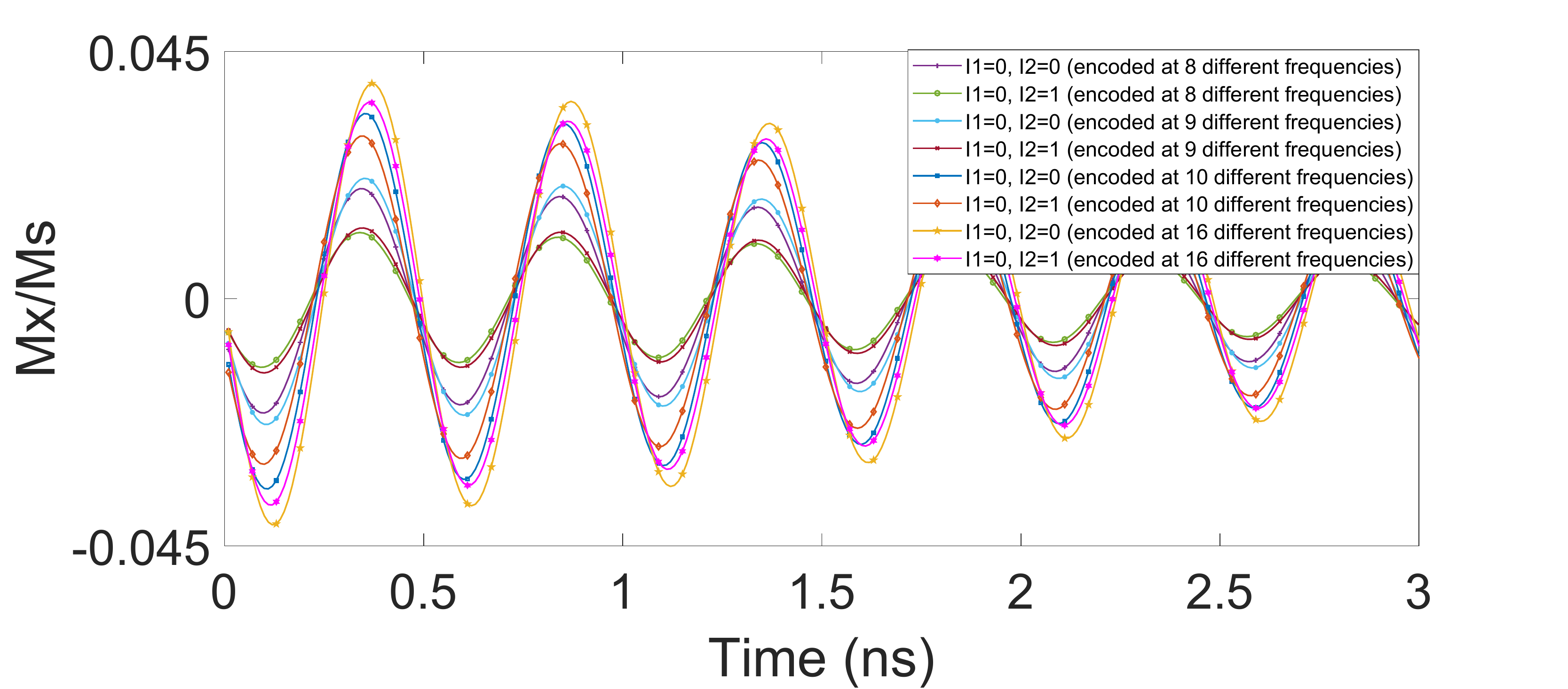}
  \caption{XOR Gate Outputs at $f_2$=$20$GHz.}
  \label{fig:results10}
\end{figure}

However, one can go beyond this limit if threshold detection based it utilized. To examine the effect of embedding more than $8$ frequencies  we evaluate by means of OOMMF simulations $2$-input XOR gates with $8$, $9$, $10$, and $16$ frequencies. For illustration purpose we display  in Figure \ref{fig:results10} the $20$ GHz frequency output component for the input combinations ${\cal I}_1{\cal I}_2=00$ and  ${\cal I}_1{\cal I}_2=01$, which should give a $0$ and $1$ output value, respectively, for all the considered input widths. 
The Figure clearly indicates that while the spin wave magnetization difference between the two input combinations decreases as the number of frequency increases, which makes output detection more challenging, two different levels can still be distinguished and a threshold defined, as such if the spin wave magnetization is greater than that threshold, the output is $0$, and $1$ otherwise. To clarify this let us inspect the output value at time moment \SI{0.4}{ns} for the $8$, $9$, $10$, and $16$-bit XOR gates. For the input combination ${\cal I}_1{\cal I}_2=00$ the output SW has a higher amplitude than the one corresponding to ${\cal I}_1{\cal I}_2=01$, which means that a threshold can be set and based on threshold detection, X(N)OR can be detected. This suggests that for threshold detection based gates are more robust and can operate with up to $16$-bit inputs. Note that more than $16$-bit inputs might be realizable  but it is part of planned future work.

\begin{figure}[t]
\centering
  \includegraphics[width=0.6\linewidth]{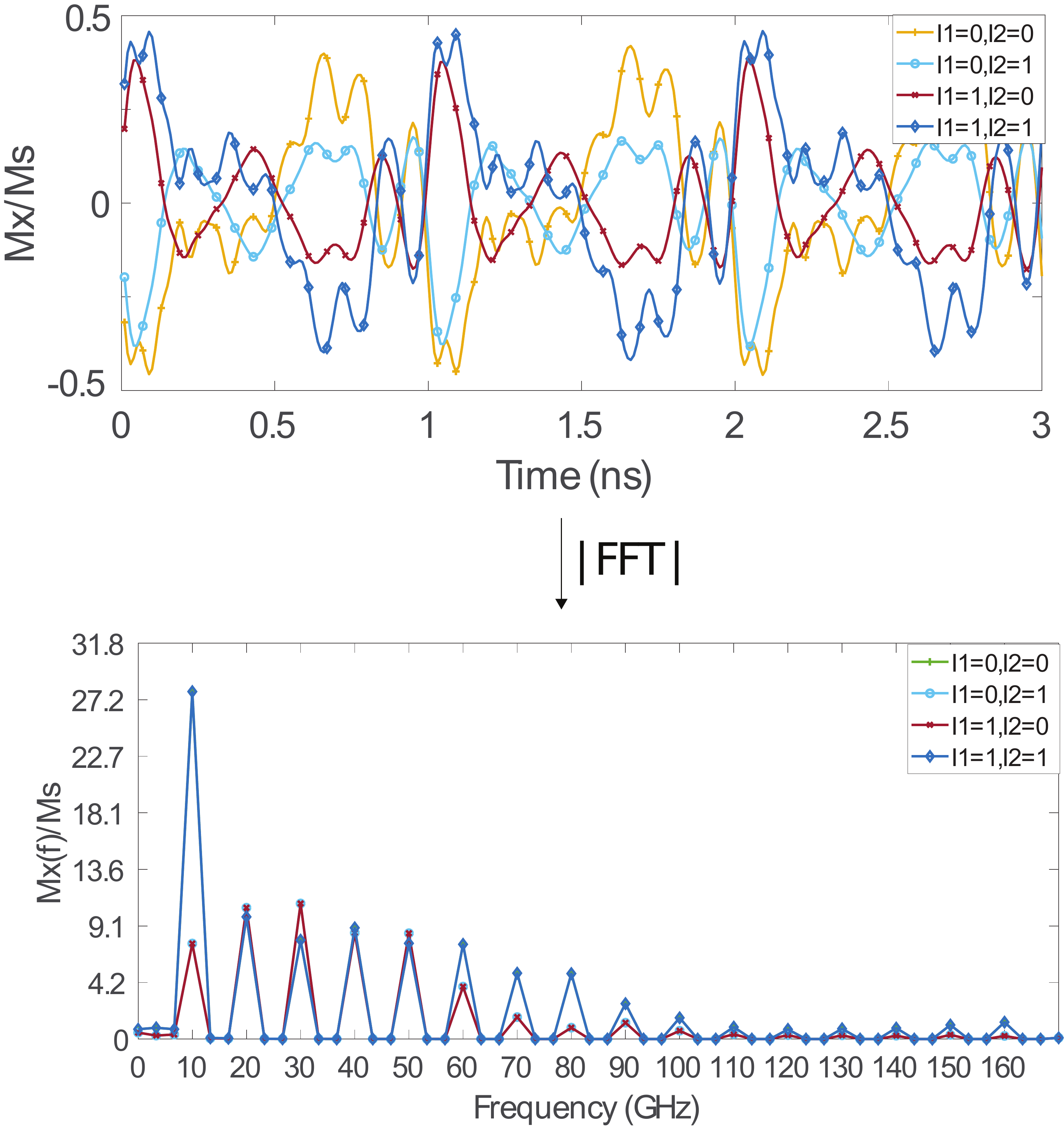}
  \caption{Optimized $16$-bit Majority Gate Response in Time and Frequency.}
  \label{fig:results11}
\end{figure} 

\begin{figure}[t]
\centering
  \includegraphics[width=0.6\linewidth]{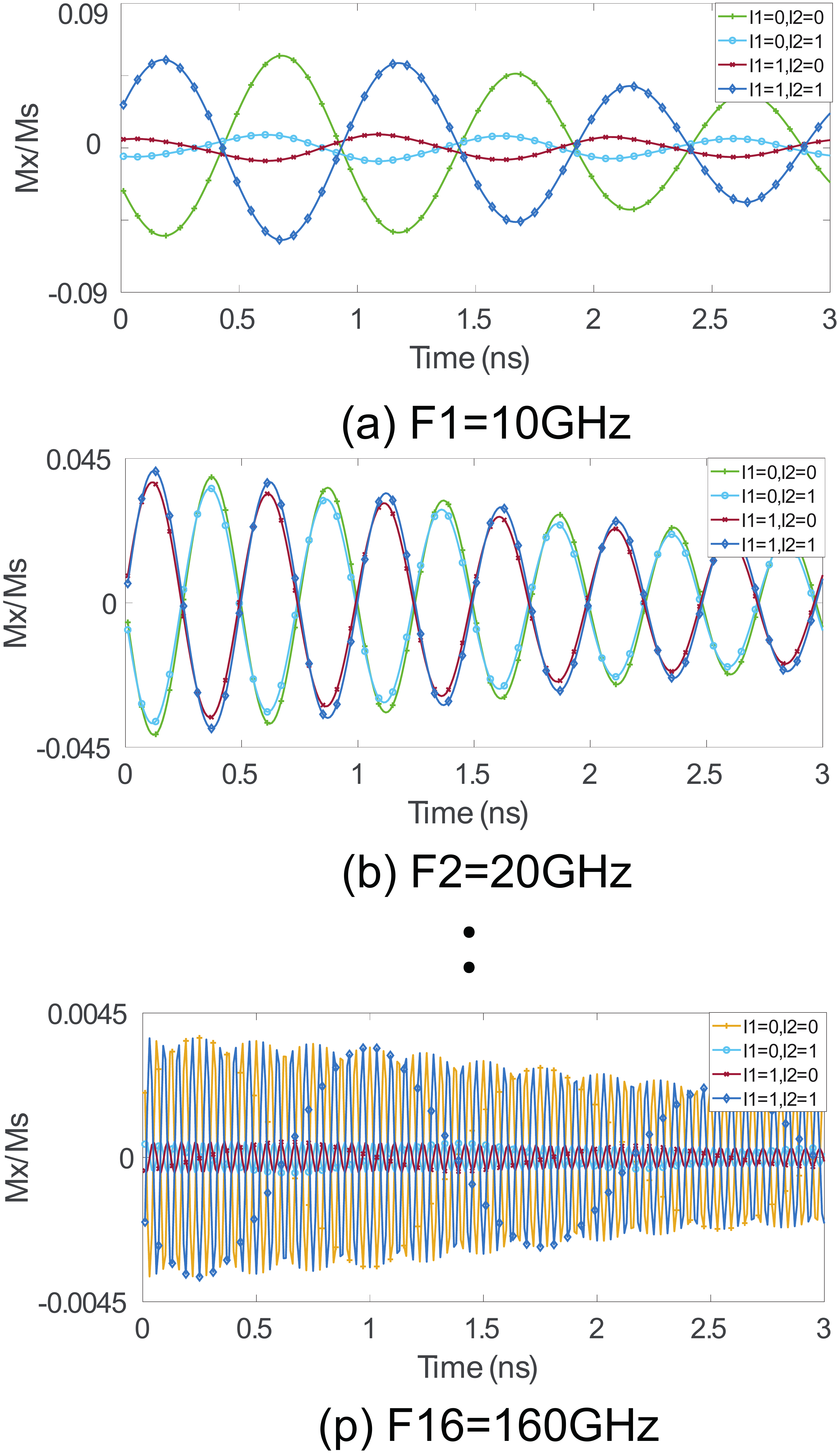}
  \caption{Optimized XOR Gate Outputs: a) $f_1$=\SI{10}{GHz}, b) $f_2$=\SI{20}{GHz}, \ldots, p) $f_{16}$=\SI{160}{GHz}.}
  \label{fig:results12}
\end{figure} 

Figure \ref{fig:results11} presents OOMMF simulation results for the $16$-bit based $2$-input XOR gate. As it can be observed from the FFT magnitude spectrum in Figure \ref{fig:results11}, the information is encoded in SWs with $16$ different frequencies, $10$, $20$, \ldots, \SI{160}{GHz} and the output for all the possible input combinations $({\cal I}_1= {\cal I}_2=0), \ldots, ({\cal I}_1={\cal I}_2=1)$ can be detected at each frequency. To further examine the results, we filter each frequency component for different input combinations separately in Figure \ref{fig:results12} and one can observe that the output SWs  are not distorted and can be properly detected at each frequency, which means that the  $16$-bit XOR/XNOR gate operates correctly. Let us consider the \SI{20}{GHz} output time moment \SI{0.75}{ns} and a detection threshold value of $0.04$ $M_s$. For ${\cal I}_1={\cal I}_2=0$, or ${\cal I}_1={\cal I}_2=1$ the absolute SW magnetization value is greater than $0.04$ $M_s$ due to the constructive interference, which means $0$ logic output as it should.  For ${\cal I}_1= 0 {\cal I}_2=1$, or ${\cal I}_1=1 {\cal I}_2=0$ the absolute SW magnetization value is lower than $0.04$ $M_s$, which means a $1$ logic output as it should.  

An XNOR can be realized by flipping the condition such that a SW magnetization lower (greater) than $0.04$ $M_s$ is read as a logic $0$ ($1$).  The same line of reasoning can be utilized to determine all threshold values as, $0.045$ $M_s$, $0.04$ $M_s$, $0.038$ $M_s$, $0.033$ $M_s$, $0.032$ $M_s$, $0.03$ $M_s$, $0.028$ $M_s$, $0.025$ $M_s$, $0.02$ $M_s$, $0.015$ $M_s$, $0.01$ $M_s$, $0.007$ $M_s$, $0.0068$ $M_s$, $0.005$ $M_s$, $0.0045$ $M_s$, $0.004$ $M_s$, $0.0035$ $M_s$, and $0.002$ $M_s$, for value increasingly ordered frequencies.

\subsection{Variability and Thermal Noise Effects}
In this paper, our main purpose is to propose and validate an intrinsic data parallel spin wave technology under ideal conditions as a proof of  concept, while disregarding  factors, e.g., edge roughness, waveguide dimension variations, spin wave strength variation, and thermal noise, which might negatively affect the performance of the proposed concept. However, in \cite{DC,DC9}, the effects of waveguide trapezoidal cross section and edge roughness were investigated and demonstrated that they have a rather limitted impact in gate behavior, which preserve functionality under their presence. Moreover, an investigation of a SW gate behaviour at different temperatures was presented in \cite{DC}. At different temperatures, it was noticed that the gate functions correctly and that the temperature variation effect is rather limited. In addition to that, as our proposed structure is in-line waveguide width variations do not affect gate functionality, thus we expected it to be rather robust to dimension variations. 
Despite that fact that we expect that variability and thermal noise do not fundamentally  affect the proposed gate behaviour, a thorough investigation of such effects is part of the planned future work.

\section{Conclusions}
\label{sec:Conclusion}

A novel $n$-bit data parallel spin wave logic gate was proposed in this paper. In order to explain the proposed concept, we implemented and validated by means of OOMMF, $8$-bit $2$-input XOR and $3$-input Majority gates. Further, we proposed an optimization algorithm to minimize the area overhead of the proposed multi-frequency gates and demonstrate that the algorithm diminishes the area by $30$\% and $41$\% for XOR and MAJ gates implementations, respectively. Moreover, to asses the potential of our proposal, we evaluated and compared the proposed multifrequency gates with functionally equivalent scalar SW gate based implementations in terms of area, delay, and power consumption. The results indicated that the byte-based XOR and Majority gates require $4.47$x and $4.16$x area less than the conventional (scalar) implementations, respectively, at the expense of $5$\% to $7$\% delay overhead and without inducing any power consumption overhead. Finally, we demonstrated that, for current gate topology and materials, the maximum number of frequencies (gate parallelism) is $8$ and $16$ for phase and  threshold based output detection, respectively. 

\section*{Acknowledgement}
This work has received funding from the European Union's Horizon 2020 research and innovation program within the FET-OPEN project CHIRON under grant agreement No. 801055. It has also been partially supported by imec's industrial affiliate program on beyond-CMOS logic. F.V. acknowledges financial support from Flanders Research Foundation (FWO) through grant No.~1S05719N.

\bibliography{Multi-frequency_Data_Parallel_Spin_Wave_Logic_Gates}

\end{document}